\documentclass
[
  aip,
  jcp,
  amsmath,
  amssymb,
  preprint,
]
{revtex4-1}
\draft
\usepackage{amsmath,mathtools, amssymb}
\usepackage{graphicx}
\usepackage{dcolumn}
\usepackage[utf8]{inputenc} 
\usepackage{bm}
\usepackage[T1]{fontenc}
\usepackage{mathptmx}
\usepackage{etoolbox}
\usepackage{hyphenat}
\usepackage{xcolor}
\usepackage[binary-units]{siunitx}
\usepackage[colorlinks, linkcolor = blue, citecolor = blue, filecolor = black, urlcolor = blue]{hyperref}
\usepackage{glossaries}
\usepackage{blkarray, bigstrut}
\usepackage{braket}
\usepackage{cleveref}
\usepackage[caption=false]{subfig}
\usepackage{chemformula} 
\usepackage[normalem]{ulem}
\usepackage{physics}
\usepackage{textcomp}
\usepackage{xr}

\newif\iflocal
\localfalse
\iflocal \input{local_header} \fi

\newacronym{isf}{iSF}{intramolecular singlet fission}
\newacronym{xsf}{xSF}{intermolecular singlet fission}
\newacronym{sf}{SF}{singlet fission}
\newacronym[firstplural = locally excited]{le}{LE}{locally excited}
\newacronym[firstplural = charge transfers]{ct}{CT}{charge transfer}
\newacronym{tt}{TT}{triplet pair}
\newacronym{pp}{PP}{projected\hyp purification}
\newacronym{dmrg}{DMRG}{density matrix renormalization group}
\newacronym{tdvp}{TDVP}{time dependent variational principle}
\newacronym[firstplural = reduced density matrices]{rdm}{RDM}{reduced density matrix}
\newacronym{meg}{MEG}{multiple exciton generation}
\newacronym{me}{ME}{multi excited}
\newacronym{casscf}{CASSCF}{complete active space self consistent field}
\newacronym{sa-casscf}{SA\hyp CASSCF}{state averaged complete active space self consistent field}
\newacronym{casci}{CASCI}{complete active space configuration interaction}
\newacronym{homo}{HOMO}{highest occupied molecular orbital}
\newacronym{lumo}{LUMO}{lowest unoccupied molecular orbital}
\newacronym{mps}{MPS}{matrix\hyp product state}
\newacronym{ttns}{TTNS}{tree tensor\hyp network state}
\newacronym{mo}{MO}{molecular orbital}
\newacronym{mctdh}{MCTDH}{multi\hyp configurational time\hyp dependent Hartree}
\newacronym{svd}{SVD}{singular value decomposition}
\newacronym{lbo}{LBO}{local basis optimization}
\newacronym{ps}{PS}{pseudo\hyp site}
\newacronym{tdwpd}{TDWPD}{time-dependent wave packet diffusion}
\newacronym{dp-mes}{DP\hyp Mes}{13,13'-bis(mesityl)-6,6'-dipentacenyl}
\newacronym{dt-mes}{DT\hyp Mes}{orthogonal tetracene dimer}

\makeatletter
\def\@email#1#2{%
 \endgroup
 \patchcmd{\titleblock@produce}
  {\frontmatter@RRAPformat}
  {\frontmatter@RRAPformat{\produce@RRAP{*#1\href{mailto:#2}{#2}}}\frontmatter@RRAPformat}
  {}{}
}%
\makeatother

\newcommand*{\nodagger}{{\vphantom{\dagger}}}
\newcommand*{\noprime}{{\vphantom{\prime}}}
\newcommand*{\imagunit}{\mathrm{i}}
\newcommand*{\enumber}{\mathrm{e}}

\newcommand*{\LandauO}[1]{\mathrm{O}\left(#1\right)}

\newcommand*{\suppref}[1]{\cref{#1}}
\newcommand*{\phonnumber}{n_\mathrm{ph}}
\newcommand*{\prettyintegral}[4]{\int\limits^{#1}_{#2}\! \mathrm{d}{#3}\, #4}
\newcommand*{\tetracenename}{1,4-bis(11-phenyltetracen-5-yl)benzene}

\definecolor{colorA}{named}{blue}
\definecolor{colorB}{named}{black}
\definecolor{colorC}{named}{red}
\definecolor{colorD}{named}{green}
\definecolor{colorE}{named}{orange}
\definecolor{colorF}{named}{purple}
\definecolor{colorG}{named}{magenta}
\definecolor{colorH}{named}{cyan}

\crefname{figure}{Figure}{Figures}
\crefname{table}{Table}{Tables}
\crefname{equation}{Equation}{Equations}

\begin{document}
\preprint{DOI: 10.1063/5.0068292}
\title{Quantum dynamics simulation of intramolecular singlet fission in covalently linked tetracene dimer}

\author{Sam Mardazad}
\email{sam.mardazad@physik.uni-muenchen.de}
\affiliation{Department of Physics, Arnold Sommerfeld Center of Theoretical Physics, University of Munich, Theresienstrasse 37, 80333 Munich, Germany}
\altaffiliation{Munich Center for Quantum Science and Technology (MCQST), Schellingstrasse 4, 80799 M\"{u}nchen, Germany}
\author{Yihe Xu}
\affiliation{School of Chemistry and Chemical Engineering, Nanjing University,  Nanjing 210023, China}
\author{Xuexiao Yang}
\affiliation{School of Chemistry and Chemical Engineering, Nanjing University,  Nanjing 210023, China}
\author{Martin Grundner}
\affiliation{Department of Physics, Arnold Sommerfeld Center of Theoretical Physics, University of Munich, Theresienstrasse 37, 80333 Munich, Germany}
\altaffiliation{Munich Center for Quantum Science and Technology (MCQST), Schellingstrasse 4, 80799 M\"{u}nchen, Germany}
\author{Ulrich Schollw\"ock}
\affiliation{Department of Physics, Arnold Sommerfeld Center of Theoretical Physics, University of Munich, Theresienstrasse 37, 80333 Munich, Germany}
\altaffiliation{Munich Center for Quantum Science and Technology (MCQST), Schellingstrasse 4, 80799 M\"{u}nchen, Germany}
\author{Haibo Ma}
\affiliation{School of Chemistry and Chemical Engineering, Nanjing University,  Nanjing 210023, China}
\email{haibo@nju.edu.cn}
\author{Sebastian Paeckel}
\affiliation{Department of Physics, Arnold Sommerfeld Center of Theoretical Physics, University of Munich, Theresienstrasse 37, 80333 Munich, Germany}
\altaffiliation{Munich Center for Quantum Science and Technology (MCQST), Schellingstrasse 4, 80799 M\"{u}nchen, Germany}
\email{sebastian.paeckel@physik.uni-muenchen.de}
\date{\today}

\begin{abstract}
  In this work we study \acrlong{sf} in tetracene para\hyp dimers, covalently linked by a phenyl group.
  In contrast to most previous studies, we account for the full quantum dynamics of the combined excitonic and vibrational system.
  For our simulations, we choose a numerically unbiased representation of the molecule's wave function, enabling us to compare with experiments, exhibiting good agreement.
  Having access to the full wave function allows us to study in detail the post\hyp quench dynamics of the excitons.
  Here, one of our main findings is the identification of a time scale $t_0 \approx \SI{35}{\femto\second}$ dominated by coherent dynamics.
  It is within this time scale that the larger fraction of the \acrlong{sf} yield is generated.
  We also report on a reduced number of phononic modes that play a crucial role in the energy transfer between excitonic and vibrational systems.
  Notably, the oscillation frequency of these modes coincides with the observed electronic coherence time $t_0$.
  We extend our investigations by also studying the dependency of the dynamics on the excitonic energy levels that, for instance, can be experimentally tuned by means of the solvent polarity.
  Here, our findings indicate that the \acrlong{sf} yield can be doubled while the electronic coherence time $t_0$ is mainly unaffected.
\end{abstract}
\maketitle
\section{Introduction}
\begin{figure}[!h]
  \centering
  \subfloat[\label{fig:tetracene-1}]{
	\includegraphics[width=.65\textwidth]{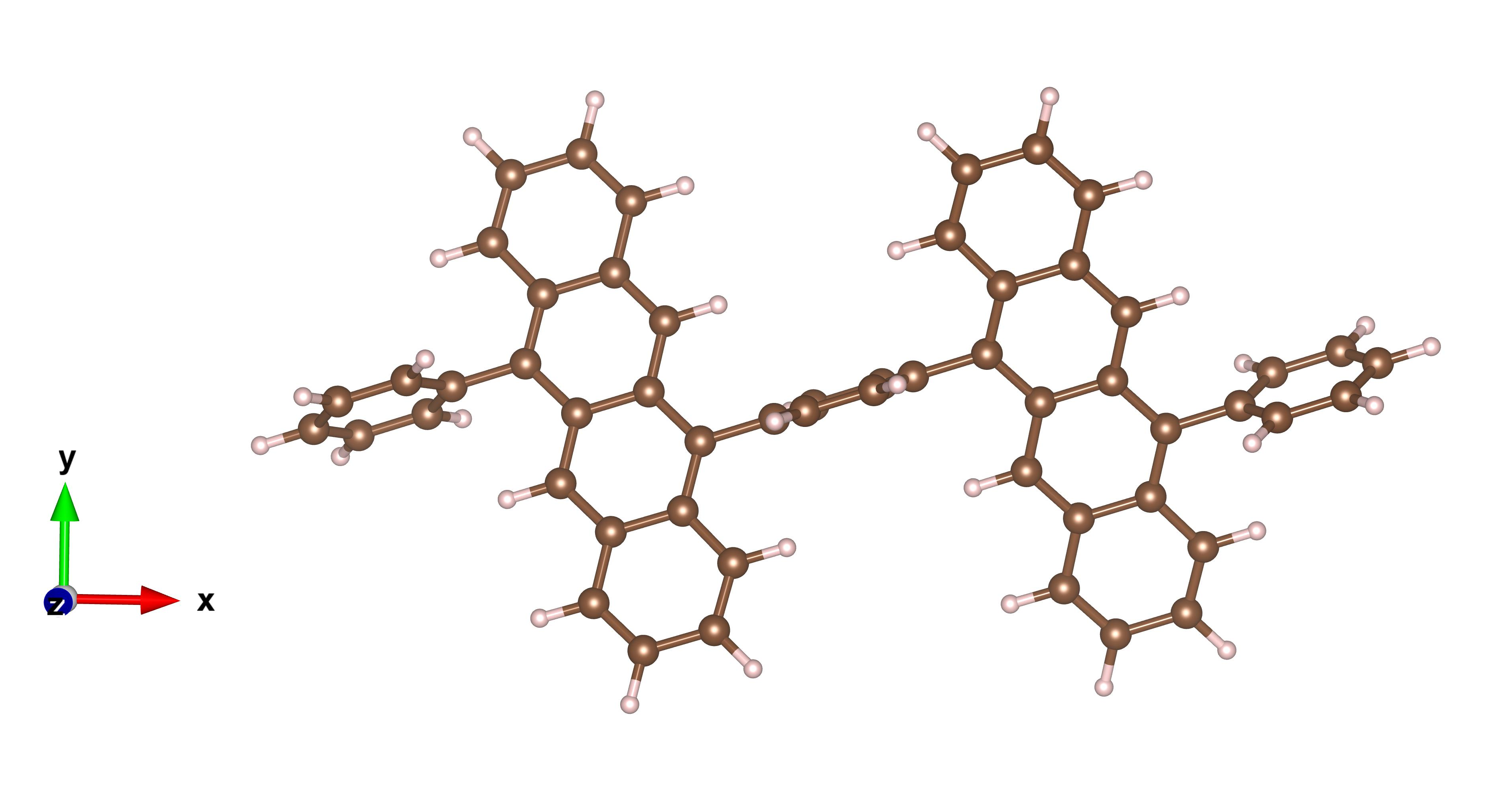}
  }%
  \subfloat[\label{fig:tetracene-2}]{
	\includegraphics[width=.25\textwidth]{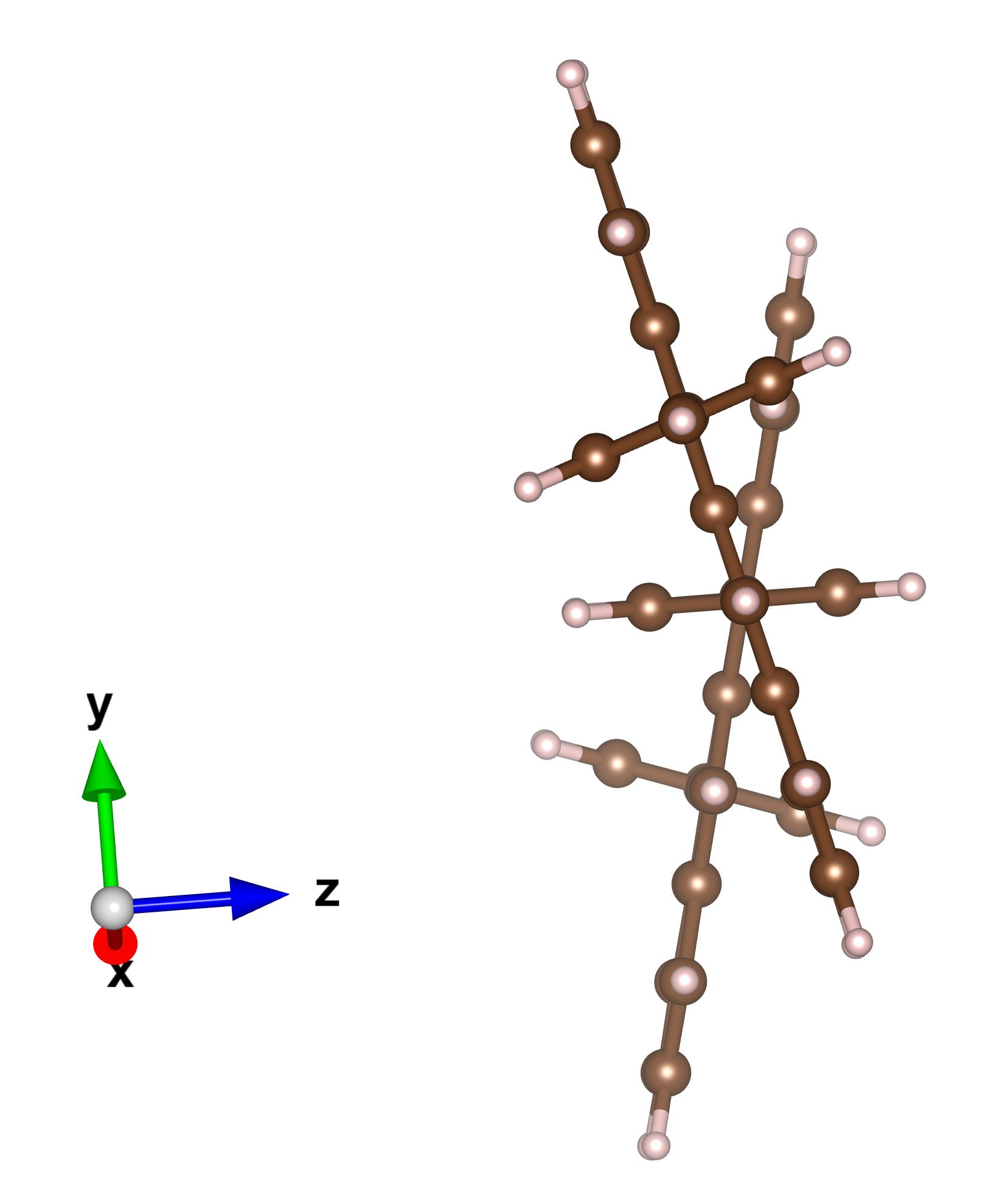}
  }

  \subfloat[\label{fig:tetracene-3}]{
	\includegraphics[width=.9\textwidth]{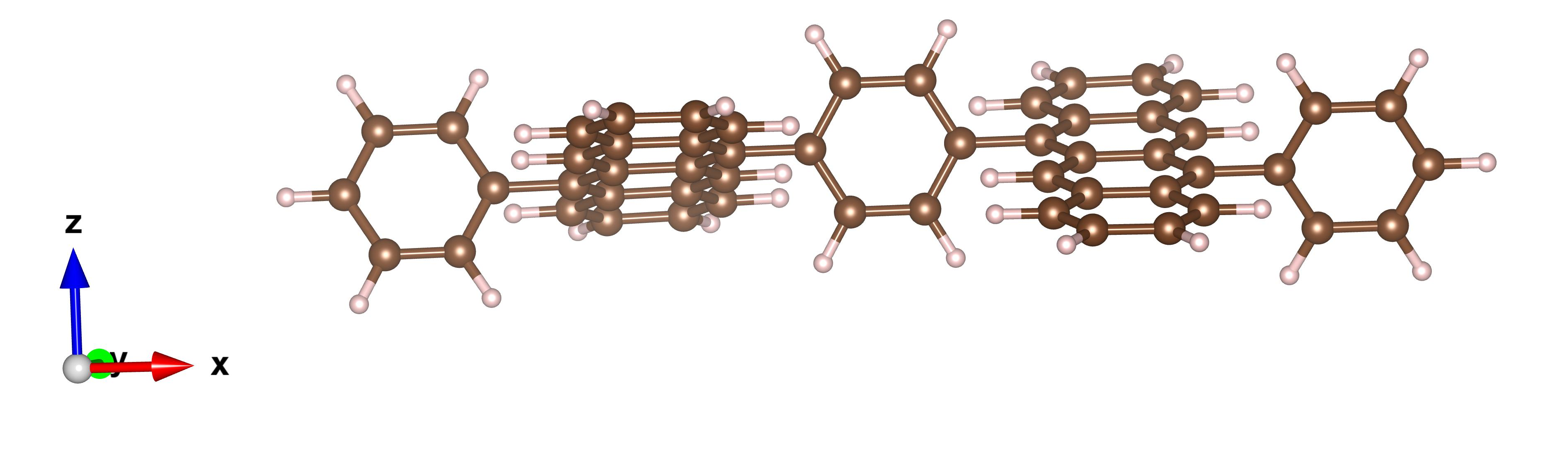}
  }
  \caption{\label{fig:tetracene_dimer_struct}\protect\cref{fig:tetracene-1,fig:tetracene-2,fig:tetracene-3} show different views of the molecular structure of~\tetracenename.}
\end{figure}

\begin{quotation}
\Gls{sf} is a spin-allowed photophysical process that generates two triplet excitons from one singlet excited state\cite{doi:10.1021/acs.jpclett.7b00230,2010SFReview,2018SFReview,zimmerman2010}.
It can realize \gls{meg} and provide longer\hyp ranged exciton transport inside a semiconductor (typically of the order of $\si{\micro\meter}$ for triplets compared to $\sim\SI{10}{\nano\meter}$ for singlets \cite{Najafov2010Nature, PhysRevLett.107.017402}).
This process gives rise to an increase in the charge\hyp carrier to radiation ratio, resulting in a net increase in the efficiency for organic semi\hyp conductor based photo\hyp voltaics.
\gls{sf} has therefore shown the potential to surpass the upper bound of the solar cell efficiency of $\SI{33}{\percent}$, set by the Shockley\hyp Queisser limit and, hence, has attracted great attention~\cite{Congreve334,doi:10.1063/1.2356795}.
\end{quotation}
Previous theoretical and experimental studies \cite{wang_isf_tetoli,2015PRLJAP,2016LZG,2018JPCLPent,2013AccountsSF,2019TGHJCP,2017JHLPCL,doi:10.1021/acs.jctc.9b00122,2015PRLJAP,monahannature,wcms.1539,doi:10.1021/acs.jctc.1c00158,doi:10.1063/5.0031435,D1CP00563D} investigated the mechanism behind \gls{sf} and distinguished between coherently driven and thermally activated \gls{sf}, where only the latter exhibits a strong temperature dependence.
The different mechanisms happen on different time scales namely up to $\SI{700}{\femto\second}$ for coherently driven \gls{sf}~\cite{monahannature} and $>\SI{1}{\pico\second}$ for thermally activated \gls{sf}~\cite{2015PRLJAP}.
Unlike widely studied \gls{xsf} systems of crystalline tetracene and pentacene, covalently linked chromophore dimers allow much easier control of inter\hyp chromophore orientation and interaction for \gls{isf}\cite{2015NMBus,2015PRLJAP}.
This is achieved by the chemical modification of the bridging group and tuning of the solvent environment.
For example, Wang et al. recently reported fast \gls{isf} within $\SI{10}{\pico\second}$ in \tetracenename , as is shown in~\cref{fig:tetracene_dimer_struct}, and illustrated its temperature dependence and significant solvent polarity effects~\cite{wang_isf_tetoli}.
\par
It is nowadays well-known that a pure electronic description is insufficient to describe these processes correctly~\cite{2017NCJAP,Bakulin2016Nature,doi:10.1063/1.4982362,2010SFReview,2018SFReview}, suggesting the necessity of the incorporation of molecular vibrations into theoretical considerations.
The main obstacles for realistic calculations are, besides others, a faithful microscopic modeling of the coupling between electronic states and vibrational degrees of freedom, and a full quantum mechanical treatment of their coupled dynamics.
Previously, due to the lack of knowledge of the full coupling Hamiltonian, the first problem has been addressed by investigating the effect of successive elimination of vibrational degrees of freedom in order to identify the relevant modes~\cite{2016LZG,2018JPCLPent}.
To accomplish a faithful simulation of the quantum dynamics following the light excitation, a variety of methods have been applied, including \gls{mctdh}~\cite{doi:10.1063/1.463007,MEYER199073,BECK20001}, Redfield theory~\cite{REDFIELD19651,MRT2003} and \gls{tdwpd}~\cite{2015ZYTDWPDRev}.
However, despite numerous investigations, there have been few numerically unbiased simulations of a full microscopic Hamiltonian that is generated from first principles for both, electronic molecular and vibronic degrees of freedom \cite{2018JPCLPent,Thoss2019isf}.
Common time\hyp evolution schemes, e.g. Redfield theory, are only valid in the weak\hyp coupling regime.
To overcome such limitations, a decomposition of the vibrational degrees of freedoms into clusters~\cite{2016LZG,2018JPCLPent}, e.g., multi\hyp layer\hyp\gls{mctdh}~\cite{Manthe2008,Manthe2009}, can be introduced, which is superficially related to the tensor\hyp network approach presented in this work.
\par
Recently, Chin, et al. applied \glspl{ttns}~\cite{Schroeder2019} to model \gls{xsf} in \gls{dp-mes} dimers.
Exploring an involved numerical procedure, they effectively computed the non\hyp perturbative real\hyp time dynamics of exponentially large vibronic wave functions of real molecules.
In this work, we attempt to adopt another novel one\hyp dimensional tensor network method to simulate \gls{isf} dynamics in realistic chemical systems.
Thereby, our main goal is to introduce a conceptually simple yet powerful and numerically well\hyp controlled procedure, as described in the following.
First, we compute the spectral density from ab initio multireference quantum chemical calculations and determine the effect of the molecule's vibrational modes on the electronic overlap integrals to obtain an effective model, capturing the coupling between electronic states and vibrations.
Employing the resulting microscopic model, we perform a simulation of the full post photo excitation quantum dynamics using a novel method from tensor networks~\cite{largelocalHilberspaces}.
In particular, this method allows for the description of many bosonic degrees of freedom with a large number of internal states by dynamically selecting the relevant bosonic modes.
The resulting, advanced numerical simulations enable us to achieve good agreement with experimental data measured for the tetracene para\hyp dimer, covalently linked by a phenyl group.
Furthermore, we are able to unambiguously identify the relevant vibrational modes dominating the post\hyp excitation dynamics by studying the energy transfer within the molecule.
Our findings of the coherent and incoherent dynamics' time scales are in very good agreement with previous investigations~\cite{2016LZG,2018JPCLPent}.
However, here, these results are based on a realistic microscopic modeling of the full molecule's dynamics.
Having access to the molecule's full wave function, we study the connection between energy transfer into the vibrational system and phonon\hyp assisted, coherently driven \gls{isf}.
Our analysis reveals a competition between an initial coherent dynamics generated by the vibrational system and its suppression due to the formation of heavy exciton\hyp phonon quasi particles.
Finally, being able to perform realistic simulations, we tuned experimentally relevant parameters such as the solvent polarity to identify the parameter regime generating the largest \gls{sf} yield.
\par
The scope of this paper is as follows:
In~\cref{sec:model}, we introduce a Hamiltonian for exciton\hyp phonon mixtures describing our system adequately.
This employs a global coupling of non\hyp local vibrations to excitonic degrees of freedom.
In~\cref{sec:numerics}, we briefly sketch the new method and exhibit performance advantages.
Furthermore, we are going to discuss the necessity for large local Hilbert spaces in simulating these types of models.
Accounting for the rapid development of tensor network techniques for large Hilbert space sizes, we also briefly elaborate on why standard, widely used methods may not be appropriate here and discuss our approach in the context of the \gls{ttns} formulation introduced by Chin et al.~\cite{Schroeder2019}.
Finally, we present the simulated \gls{isf} dynamics in~\cref{sec:results}.
These involve comparison to experiment, as well as detecting the phononic modes dominating the energy transfer.
We find a total triplet population between $\SI{14}{\percent}$ and $\SI{25}{\percent}$ depending on the solvent, comparable to the order of magnitude of experimental findings, i.e., $\SI{21}{\percent}$~\cite{wang_isf_tetoli}.
Eventually, we investigate the tuning of experimental parameters, i.e. the solvent polarity, on the total triplet yield.
\section{Model}\label{sec:model}
To model the \gls{sf} process, we describe the diabatic electronic states coupled to vibrational modes by means of a Frenkel\hyp exciton\hyp Hamiltonian~\cite{PhysRev.37.17,PhysRev.37.1276},
\begin{align}
  \label{eq:Hamiltonian}
  \hat{H} &= \sum_{ij} V_{ij} \hat{c}^\dagger_i \hat{c}^\nodagger_j + \sum_I \omega_I \hat{b}^\dagger_I \hat{b}^\nodagger_I + \frac{1}{\sqrt{2}} \sum_{ij,I} g_{ij,I} \left( \hat{b}^\dagger_I + \hat{b}^\nodagger_I \right) \hat{c}^\dagger_i \hat{c}^\nodagger_j\; \text{,}
\end{align}
where the lower\hyp case indices $i$ and $j$ correspond to the excitonic states and the upper\hyp case indices $I$ denote the vibrational modes.
Here, $\hat{c}^{(\dagger)}_i$ correspond to the usual excitonic creation and annihilation operators, while $\hat{b}^{(\dagger)}_I$ create and annihilate phonons. 
The diagonal elements ($i = j$) of $V_{ij}$ resemble the energies of the excitonic states, while the off\hyp diagonal elements ($i \neq j$) correspond to the couplings among them.
$\omega_I$ is the frequency of the vibration mode $I$, and $g_{ij,I}$ represents the coupling between the excitonic Hamiltonian and vibrational modes $I$.
We can distinguish between diagonal and off\hyp diagonal couplings here as well.
In order to describe our system adequately, we use five diabatic states.
They are the two \gls{le} states $\text{LE}_1$ and $\text{LE}_2$, the two \gls{ct} states $\text{CT}_1$ and $\text{CT}_2$, and the \gls{tt} state. 
While the \glspl{le} model the electronic excitation in the dimer subsystems, the \glspl{ct} are energetically high lying states that, if occupied, represent intermediate cationic or anionic molecular states \cite{2019SFReview}.
By initializing the system in one of the \gls{le} states (or a coherent superposition), we model a photo\hyp excitation.
%
%
It was previously shown that the localized triplets in the chromophore units can be interpreted as an electron\hyp hole bound state~\cite{zimmerman2010}.
However, these triplets in each subsystem are correlated into an overall singlet state~\cite{2010SFReview,PhysRevB.1.896}, which we call \gls{tt}.
The dissociation of the \gls{tt} state into locally separated triplets is a topic on its own and not part of this investigation\cite{TTspin2016JCTC,2019SFReview}.
The details of the construction of these five diabatic states from \gls{sa-casscf} excited state calculations can be found in \suppref{sec:elecstruct,sec:redfieldsearch}.
The diabatic exciton matrix elements $V_{ij}$ thus obtained are shown in \cref{table:electronic_coupl}.
\begin{table}
  \centering
    \caption{\label{table:electronic_coupl}Excitonic Hamiltonian elements $V_{ij}$ and their corresponding thermal fluctuations $\sigma_{ij}$ (defined in \cref{eq:elec_Hamil_temp_fluc}) at $\SI{300}{\kelvin}$ in units of $\si{\electronvolt}$.The first entry per cell is $V$, while the second entry denotes $\sigma$ for the given pair of diabatic states $i,j$.}
  \begin{tabular}{l||rr|rr|rr|rr|rr}
    \hline
                         &
    \multicolumn{2}{r|}{$\ket{\text{LE}_1}$}  &
    \multicolumn{2}{r|}{$\ket{\text{LE}_2}$}  &                    
    \multicolumn{2}{r|}{$\ket{\text{CT}_1}$}  &
    \multicolumn{2}{r|}{$\ket{\text{CT}_2}$}  &
    \multicolumn{2}{r}{$\ket{\text{TT}}$}   
    \\
    \hline
                         
    $\ket{\text{LE}_1}$   & 
     2.9635 & 0.279     &    
     -0.0857 & 0.035    &    
     0.0431 & 0.039     &    
     -0.0535 & 0.052    &    
     0.0002 & 0.000          
    \\
    $\ket{\text{LE}_2}$ & & &
     2.9637 & 0.280    &    
     0.0535 & 0.052   &    
    -0.0431 & 0.040   &    
    -0.0002 & 0.000        
    \\
     $\ket{\text{CT}_1}$ & & & & &
     3.3645 & 0.276   &    
    -0.0007 & 0.001   &    
    -0.0612 & 0.055        
    \\
     $\ket{\text{CT}_2}$ & & & & & & &
     3.3645 & 0.276   &    
    -0.0612 & 0.055        
    \\
    $\ket{\text{TT}}$ & & & & & & & & &
     3.1493 & 0.516         
    \\
    \hline
  \end{tabular}

\end{table}
Note that the multi-exciton \gls{tt} state is optically dark~\cite{zimmerman2010,doi:10.1021/nl062059v}.
Furthermore, the coupling between \gls{le} and \gls{tt} is almost zero, and, thus direct conversion is strongly suppressed.
This originates from the fact that the direct process corresponds to a double-electron transfer.
The related two\hyp electron integrals, between \acrshortpl{homo}/\acrshortpl{lumo} of different groups or molecules, are usually vanishingly small\cite{doi:10.1021/acs.jctc.9b00122,2017JACS2ELE,2018SFReview}.
However, the couplings between \gls{ct} and  \gls{tt} or \gls{le}  are relatively large $(\SI{40}-\SI{60}{\milli\electronvolt})$.
This suggests a possible indirect superexchange \gls{isf} path (\gls{le}$\rightarrow$\gls{ct}$\rightarrow$\gls{tt})~\cite{doi:10.1063/1.4794425,doi:10.1063/1.4794427,Lukman_2016, Margulies_2016, Johnson_2013}.
It should be mentioned that due to the lack of dynamical correlation in the \glsunset{casscf}\gls{casscf} calculations, the diagonal entries of the excitonic matrix $V_{ij}$ may be inaccurate.
Therefore, we tune the \gls{le} and \gls{tt} energies to fit the peak position of experimental absorption spectra.
Furthermore, the energies of \gls{ct} are obtained from a linear search within the range of  $\SI{3.0645}-\SI{3.5645}{\electronvolt}$ using Redfield dynamic simulations~\cite{MRT2003}.
The details about this procedure can be found in~\suppref{sec:redfieldsearch}.
%
%
\par
\begin{figure}[!ht]
  \centering
   \subfloat[0.45\textwidth][\label{fig:spectral_density_lele}] {
     \iflocal \include{figures/spectral_density_le1le1} \else
     \includegraphics[width = 0.42\textwidth]{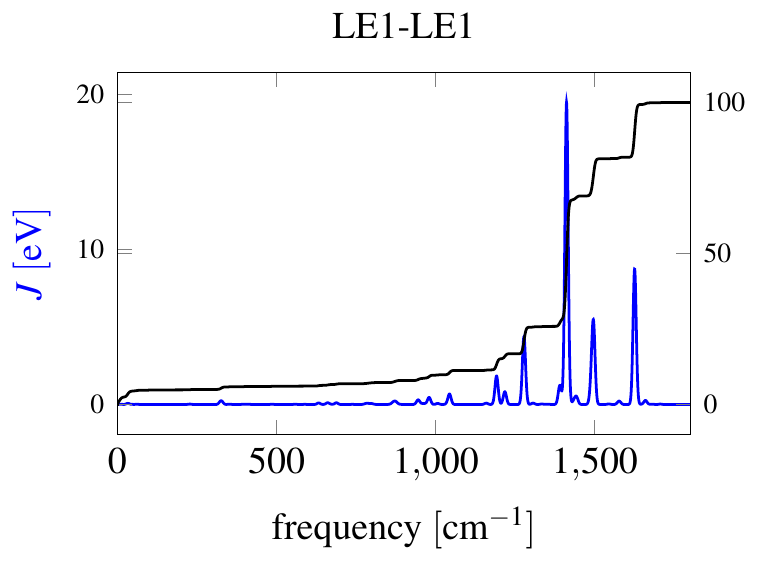} \fi
   }
   \subfloat[0.45\textwidth][\label{fig:spectral_density_lect}] {
     \iflocal \include{figures/spectral_density_le1ct1} \else
     \includegraphics[width = 0.45\textwidth]{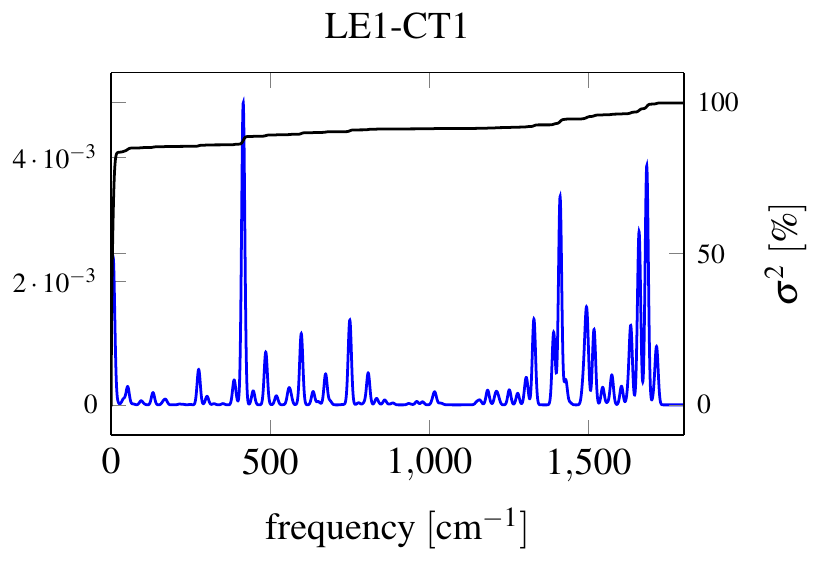} \fi
   }\\
   \subfloat[0.45\textwidth][\label{fig:spectral_density_cttt}] {
     \iflocal \include{figures/spectral_density_ct1tt} \else
     \includegraphics[width = 0.45\textwidth]{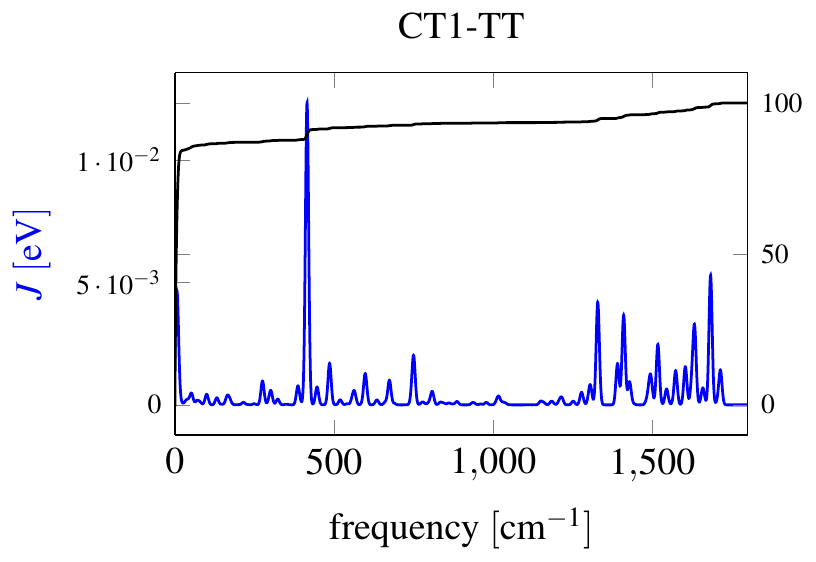} \fi
   }
   \subfloat[0.45\textwidth][\label{fig:spectral_density_tttt}] {
     \iflocal \include{figures/spectral_density_tttt} \else
     \includegraphics[width = 0.42\textwidth]{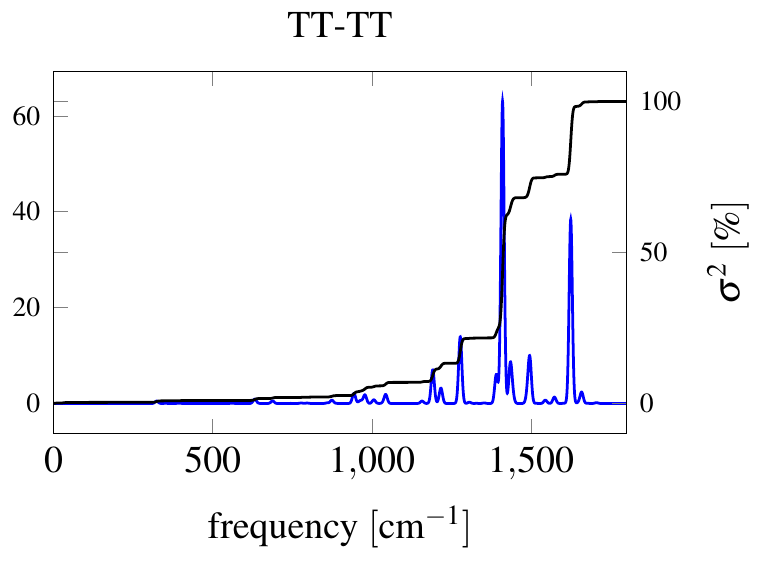} \fi
   }
  \caption{\label{fig:spectral_densities}\protect\cref{fig:spectral_density_lele,fig:spectral_density_lect,fig:spectral_density_cttt,fig:spectral_density_tttt} show the spectral density (blue, left axis) and accumulated fluctuations (black, right axis) of selected excitonic Hamiltonian terms at room temperature.}
\end{figure}

To estimate the coupling strength between excitonic states and vibrations we calculate the spectral density according to
\begin{align}
   \label{eq:spectral_density}
   J_{ij}(\omega) &= \sum_{I} g_{ij,I}^{2}\,\delta(\omega-\omega_{I}) \; ,
\end{align}
and the temperature-dependent fluctuations of the excitonic Hamiltonian elements using
\begin{align}
   \label{eq:elec_Hamil_temp_fluc}
   \sigma_{ij}^{2} &= \prettyintegral{\infty}{0}{\omega}{J_{ij}(\omega)\coth \frac{\omega}{2kT}} \; .
\end{align}
The details of the methodology for evaluating the exciton-phonon couplings and the graphical illustrations of some essential vibrational modes can be found in~\suppref{sec:vibmodes,sec:specdensity}.
The results are shown in \cref{table:electronic_coupl} and \cref{fig:spectral_densities}.
From the selected spectral density displayed in \cref{fig:spectral_densities}, we can see that for diagonal elements, there are high-frequency modes near $\SI{1400}{\per\centi\meter}$ dominating the coupling.
The remaining spectral densities for other couplings can be found in \suppref{sec:specdensity}.
Mode nos. 184 ($\SI{1409.61}{\per\centi\meter}$) and no. 185 ($\SI{1411.18}{\per\centi\meter}$) account for these peaks.
They change the backbone of the tetracene chromophore unit (see~\suppref{sec:vibmodes}).
This will influence the \gls{mo} coefficients and integrals substantially and changes the energy of diabatic states consequently.
At the same, time the lower frequency mode at $\SI{415}{\per\centi\meter}$, which mainly twists the plane of the bridging phenyl group (see~\suppref{sec:vibmodes}), mostly contributes to the coupling in off-diagonal elements.
Under this oscillation, the tetracene group remains unaltered, and therefore, the diagonal elements are less sensitive to this mode.
The off-diagonal elements, however, depend on the overlap between orbitals from different tetracene chromophore units.
The latter are dominated only to a small part by the orbitals in the bridge because the tetracene groups are far away from each other.
Consequently, twisting the bridge will change only the off-diagonal elements significantly.
Furthermore, the lowest frequency mode around $\SI{7}{\per\centi\meter}$ also contributes to off-diagonal elements because it changes the dihedral angle between the planes of two tetracene units.
\par
In order to maintain numerical feasibility we have to pick the relevant vibrational modes of the $88$ atomic molecules.
Out of the $258$ vibrational modes we neglect 181 that trigger relative fluctuations smaller than $\SI{0.1}{\percent}$.
We also discard the modes with frequencies below $\SI{10}{\per\centi\meter}$ since their oscillation periods are longer than $\SI{200}{\femto\second}$.
This time scale is set by the time\hyp range of our quantum dynamics simulation.
Finally, we couple the remaining 76 modes, ranging from frequencies $\SI{10.18}{\per\centi\meter}$ ($\SI{0.0013}{\electronvolt}$) to $\SI{1714.2}{\per\centi\meter}$ ($\SI{0.2125}{\electronvolt}$), to the excitonic system.
\section{Methods}\label{sec:numerics}
Simulating the full quantum dynamics of~\cref{eq:Hamiltonian} requires a careful and well\hyp converged treatment of the relevant vibrational modes.
For that reason, we exploit a \gls{mps} representation~\cite{Schollwoeck_2005,Schollwoeck_2011,TDVPreview} of the full quantum many\hyp body wave functions to represent the time\hyp evolution of the photo\hyp excited system.
The major challenge is the fact that each vibronic mode, in principle, requires an infinite number of internal degrees of freedom.
For practical purposes, we need to truncate the local Hilbert space dimensions to a large, but finite, number $d = n_\mathrm{ph} + 1$.
In doing so, care must be taken to choose the local dimensions large enough such that the dynamics is well\hyp behaved.
Otherwise, the system will notice the upper limit of the allowed vibrational mode occupations.
This truncation needs to be balanced with computational cost scaling as $\LandauO{d^3}$ in the local dimension for the two\hyp site \gls{tdvp} method, which we used in our calculations to perform the time evolution~\cite{Haegeman_2011,Haegeman_2016}.
The numerical calculations were carried out using the SyTen toolkit \cite{hubig:_syten_toolk}.
\par
We overcome these issues by employing the newly developed method of \gls{pp}\cite{largelocalHilberspaces,comparative}.
Within this approach, we restore the $U(1)$\hyp particle number conservation symmetry by doubling the size of the (phononic) system.
Furthermore, we introduce a dynamical procedure to truncate the local phononic Hilbert space dimensions $d$ adaptively by the choice of our truncated weight ($\delta = \num{1.e-8}$).
This enables us to treat large local Hilbert spaces with $n_\text{ph}=63$ very efficiently with the approximation being well\hyp controlled by the maximally allowed sum of discarded Schmidt values.
\subsection{Representing the vibrational modes}
\begin{figure}[!ht]
	\begin{minipage}{0.39\textwidth}
		\subfloat[\label{fig:pp-sketch}] {
			\centering
			\iflocal \input{figures/pp_sketch} \else
			\includegraphics[width=\textwidth]{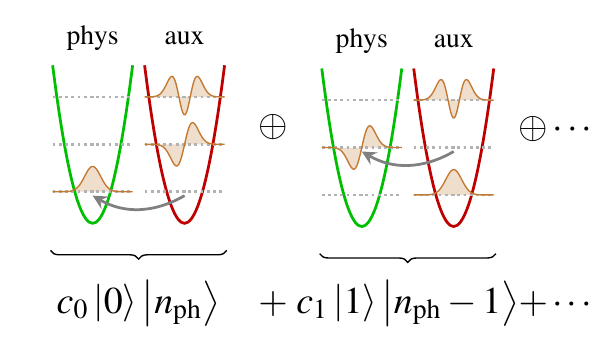} \fi
		}
		\\
		\subfloat[\label{fig:pp-geometry}] {
			\centering
			\iflocal \input{figures/pp_geometry} \else
			\includegraphics[width=\textwidth]{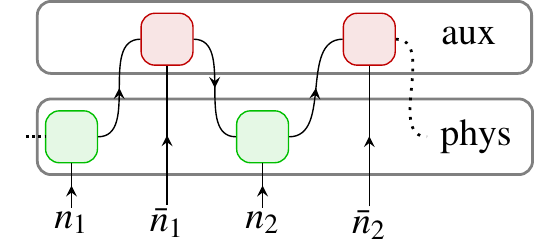} \fi
		}
	\end{minipage}%
	\hfill
	\begin{minipage}{0.6\textwidth}
		\subfloat[\label{fig:runtimes}] {
			\centering
			\iflocal \input{figures/runtimes} \else
			\includegraphics[width=0.9\textwidth]{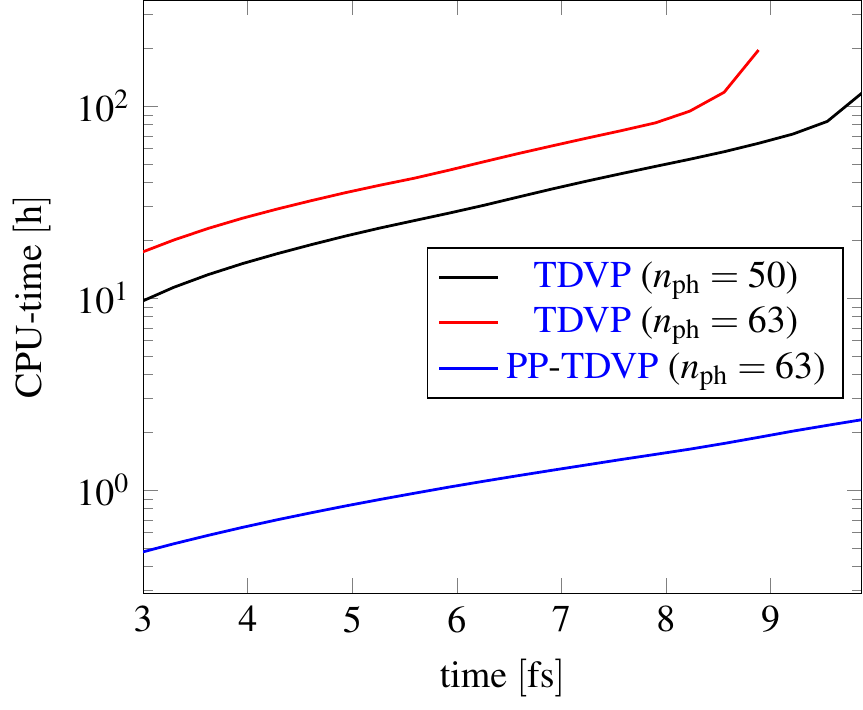} \fi
		}
	\end{minipage}
	\caption{
		\protect\subref{fig:pp-sketch} Illustrates the effect of the \gls{pp} mapping.
		Each vibronic degree of freedom is accompanied with an auxiliary site.
		Vibronic excitations are modelled by moving occupations from the auxiliary to the physical system, introducing a global $U(1)$ symmetry.
		This symmetry can be exploited by choosing a \gls{mps} representation that pairs up physical and auxiliary sites as shown in \protect\subref{fig:pp-geometry}.
		The obtained CPU times for $n_\mathrm{ph} = 50,63$ without the \gls{pp}\hyp mapping and for $n_\mathrm{ph} = 63$ with \gls{pp} are displayed in \protect\subref{fig:runtimes}.
		All calculations were carried out on an Intel\textsuperscript\textregistered Xeon\textsuperscript\textregistered Gold 6130 CPU ($\SI{2.10}{\giga\hertz}$) with 32 threads.
		Note the logscale on the y-axis.
	}
\end{figure}

Let us briefly summarize the key points of this procedure.
We start with a wave function expanded in a local basis labelled by irreducible representations of a global operator $\hat{N}$
\begin{align}
  \label{eq:originial_psi}
  \ket{\psi} &= \sum_{n_1\dots n_L} c_{n_1\dots n_L} \ket{n_1\dots n_L} \; \text{.}
\end{align}
Here, $L$ corresponds to the total number of single particle orbitals and the $n_i$ run from $\{0,\dots,d-1\}$.
Exploiting global $U(1)$ symmetries typically provide a significant speedup in numerical simulations~\cite{Singh_2011}, however, the number of phonons is not conserved in~\cref{eq:Hamiltonian}.
We circumvent this problem by introducing an auxiliary environment Hilbert space, which is a copy of the original one,
\begin{align}
  \label{eq:double_psi}
  \ket{\psi} = \sum_{\substack{n_1\dots n_L \\ \bar n_1\dots \bar n_L}} c_{n_1\dots n_L \bar n_1\dots \bar n_L} \ket{n^\noprime_1\dots n^\noprime_L} \ket{\bar n_1\dots \bar n_L} \; \text{.}
\end{align}
Here, pure indices $n^\noprime_j$ correspond to physical degrees of freedom, while those carrying a bar $\bar{n}_j$ involve the auxiliary space.
We further introduce the gauge condition
\begin{align}
  c_{n_1\dots n_L \bar n_1\dots \bar n_L} = 0 \qquad \Leftrightarrow \qquad \forall n_i + \bar n_i \neq n_\mathrm{ph} = d-1 \; \text{.}
\end{align}
This transforms \cref{eq:double_psi} into
\begin{align}
  \label{eq:purified_psi}
  \ket{\psi} &= \sum_{n^\noprime_1\dots n_L} c_{n_1\dots n_L, n_\mathrm{ph}} \ket{n_1\dots n_L} \ket{n_\mathrm{ph}- n_1,\dots ,n_\mathrm{ph}- n_L} \; \text{,}
\end{align}
which happens to be an eigenstate of $\hat{N} + \hat{\bar N}$, i.e., the total particle number operator expanded to the enlarged Hilbert space, with particle number $Ln_\mathrm{ph}$.
\Cref{fig:pp-sketch} illustrates this idea, showing the effect of the gauge constraints at the example of the two lowest excitations of a vibrational mode.
It can be shown~\cite{largelocalHilberspaces} that there exists a bijective mapping between the basis in the original Hilbert space and the basis of the doubled one.
Note that the subspace spanned by all basis states $\left\{\ket{n_1\dots n_L} \ket{n_\mathrm{ph}- n_1\dots n_\mathrm{ph}- n_L}\right\}_{n_1,\ldots,n_L}$ has the same dimension as the original one.
We also map arbitrary global operators to the purified Hilbert space, by substituting creation and annihilation operators with
\begin{align}
  \hat b_i &= \sum_{n=0}^{n_\mathrm{ph}-2} \sqrt{n+1} \ket{n}_i\prescript{}{i}{\bra{n+1}} \otimes \ket{n+1}_i \prescript{}{i}{\bra{n}}  \\
  \hat b^\dagger_i &= \sum_{n=0}^{n_\mathrm{ph}-2} \sqrt{n+1} \ket{n+1}_i\prescript{}{i}{\bra{n}} \otimes \ket{n}_i \prescript{}{i}{\bra{n+1}}
\end{align}
in terms of the operators breaking the global $U(1)$ symmetry. The remaining operators are tensored with identities in the auxiliary space.
With this representation of $\hat{H}$, one always stays within the purified sub\hyp manifold~\cite{largelocalHilberspaces}.
\par
Adopting this representation, we have obtained a formal global $U(1)$ symmetry for the phonon system that is not particle number conserving, at the cost of introducing a reservoir degree of freedom for each bosonic orbital, as shown in~\cref{fig:pp-geometry}.
Instead of treating a few huge matrix blocks, we can now compute with a lot of small ones.
We exploit the potentially large number of blocks by parallelizing contractions over the symmetry blocks.
Furthermore, we make careful use of the real space parallel \gls{tdvp} algorithm~\cite{PhysRevB.101.235123} for some of the solvent dependent calculations.
\par
Importantly, the \gls{pp} mapping allows us to variationally optimize in the space of retained optimal phononic modes by the tuning of one parameter, namely, the truncated weight during the \gls{svd}.
This can be understood by means of the relation between the retained singular values $s_\tau$ during the decimation process of the \gls{dmrg}~\cite{Schollwoeck_2005} and the diagonal elements of the \gls{rdm} of a single bosonic site $\rho$~\cite{largelocalHilberspaces},
\begin{align}
  \sum_\tau \left(s^n_\tau\right)^2 &= \rho_{nn} \; \text{.}
\end{align}
Here, the index $n$ indicates the singular values belonging to the corresponding irreducible representation.
Discarding the smallest singular values fulfilling $\sum\limits_{n,\tau} \left(s^n_\tau\right)^2 \leq \delta$ corresponds to eliminating occupation number elements $\rho_{nn}$ not required for an adequate description of the state.
Hence, the initial local Hilbert space dimensions $d$ of the vibronic modes are truncated to significantly smaller values, rendering the simulations numerically feasible.
During the dynamics, the truncation process of the state enforces the system to keep only the modes required for an adequate approximation of the state (i.e., with sufficient weight).
\par
The restored global $U(1)$ symmetry of the phononic system combined with this truncation scheme results in a highly diminished memory requirement and, more importantly, reduction in CPU time by a factor of 100-1000, due to decomposition and parallelization respectively, as is shown in~\cref{fig:runtimes}.
We compare the total CPU times during the initial stage of the dynamics, following a photo excitation of a bright exciton with and without exploiting the \gls{pp} mapping and for different local Hilbert space dimensions $d$ in the case of the trivial phonon representations.
Note that in the conventional approach, the memory requirements grow rapidly, causing a breakdown of the simulations without the \gls{pp} mapping after times of only $\SI{8.5}{\femto\second}$ or $\SI{9.9}{\femto\second}$, respectively.
\subsection{\label{sec:numerics:stability}Numerical stability and effects of insufficiently large vibrational Hilbert spaces}
\begin{figure}[!ht]
  \centering
    \iflocal \input{figures/new_phon_rdm} \else
    \includegraphics[width=0.9\textwidth]{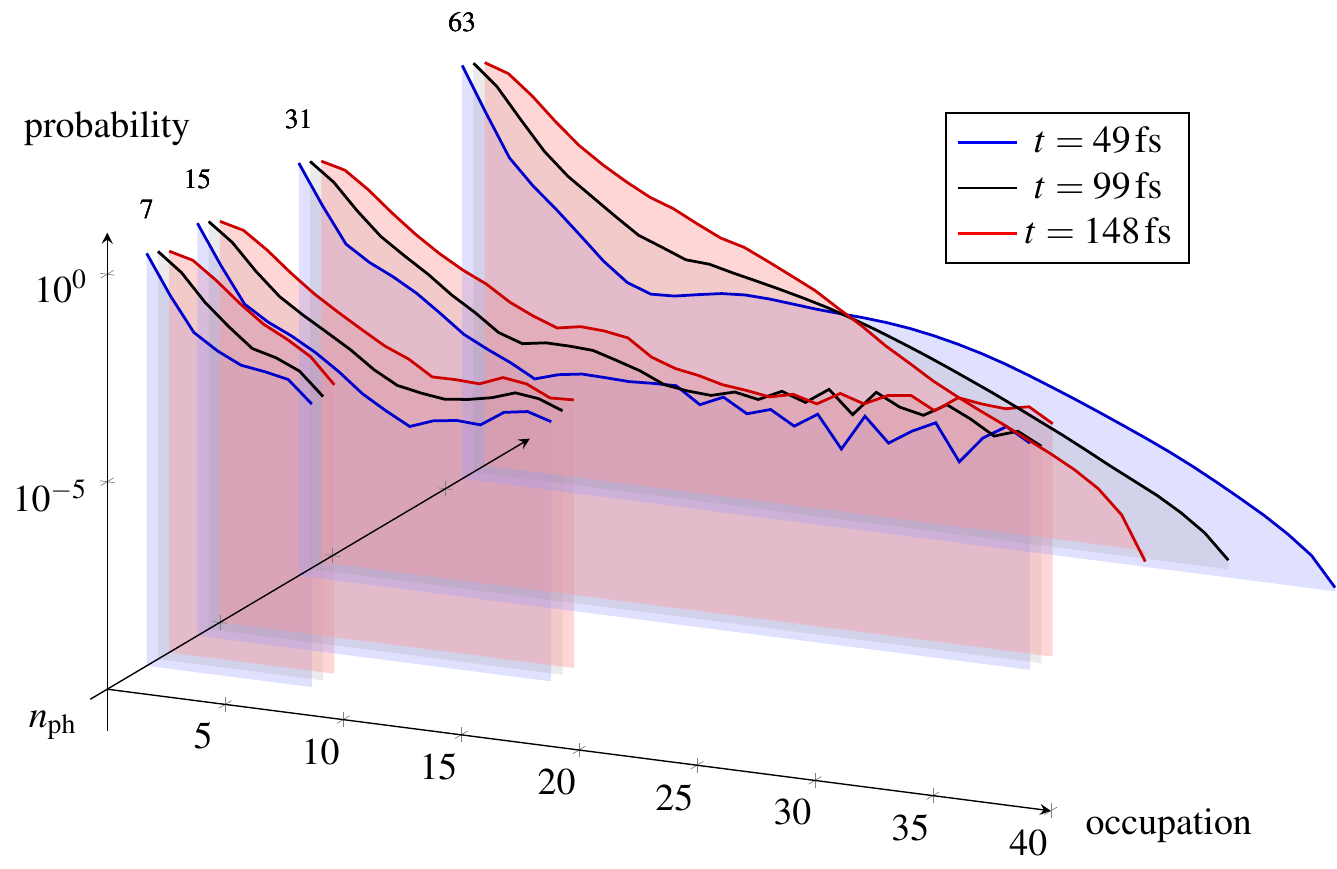} \fi
  \caption{\label{fig:phon_rdm_184}Excitation probability for occupations of phonon mode no. 184 for different times during the dynamics.}
\end{figure}
Treating the exciton\hyp phonon coupling in a numerically controlled manner is crucial in order to describe the dynamics of the vibrational modes correctly.
For that purpose, we monitor the probability of different phonon occupation numbers by calculating the \gls{rdm} during the time evolution.
The \gls{rdm} with respect to a vibronic mode $J$ is obtained by tracing out all degrees of freedom from the total density operator except for the target mode $J$,
\begin{align}
  \label{eq:phon_rdm}
  \hat{\rho}_J= \tr_{\{I\}\setminus J} \ket{\psi}\bra{\psi} \; \text{,}
\end{align}
where $\{I\}$ is the set of all phononic modes.
Exemplarily, we investigate the dynamics of a particular phonon mode (no. 184), which plays a dominant role in the excitonic dynamics (see also~\cref{fig:spectral_densities} and~\suppref{sec:specdensity}).
In order to clarify the role of the truncated local Hilbert space, we systematically tune the maximally allowed phonon numbers, i.e. $n_\mathrm{ph} = 7, 15, 31, 63$.
The result for the vibrational mode no. 184 is shown, in an exemplary fashion, in~\cref{fig:phon_rdm_184}.
It shall be mentioned that intermediate sizes, e.g. $\phonnumber = 23$, show the same behavior as the others and are therefore left out for clarity.
We evaluate the \gls{rdm} at three distinct times after photo excitation, namely, \SIlist{49;99;148}{\femto\second}.
We observe that the dynamics of the probabilities to excite only a small number of $\sim 10$ vibrational quanta is correctly described for all maximally allowed phonon numbers $n_\mathrm{ph}$.
However, inspecting the tail of the excitation probabilities, significant deviations are visible at the order of $\delta \rho_{J,nn > 10} \sim 10^{-5}$, as long as $n_\mathrm{ph} < 63$.
This translates into a failure of describing the relaxation of the initially excited phonon modes in the scope of the time evolution if the local Hilbert space dimensions are too small.
We relate this problem to the contribution of the higher excited vibrational modes to the overall time evolution operator $\mathrm{e}^{-\imagunit \hat H t}$.
Here, the vibrational energy contributions of $\hat H$ scale as $\left\langle \hat n_J \right\rangle \propto \sum\limits_n n\times \rho_{J,nn}$.
Inspecting \cref{fig:phon_rdm_184}, we find that for $n_\mathrm{ph} < 63$, the error is dominated by excitation probabilities of $\rho_{J,nn = 10-30} $, i.e., an overall error of
\begin{align}
  \sum\limits_{n=10}^{30} n \delta \rho_{J,nn} \sim 10^{-3} \; \text{.}
\end{align}
This means that the reported deviations contribute overall errors in the time evolution scaling as $\sim 10^{-3}$ in the case of mode no. 184.
\par
These observations clearly show that the higher occupation numbers of the vibrational modes can make up for non-negligible contributions to the \gls{isf} quantum dynamics, even though the mean occupation and displacement of all vibrations can be quite small.
This implies the necessity of employing a large, maximally allowed phonon number $n_\text{ph}$ per lattice site.
Throughout the simulations, the system dynamically chooses the actually required phonon Hilbert space dimension; therefore, we are able to truncate away unnecessary modes without any additional numerical effort.
\subsection{Relation to other tensor network techniques}
Finally, we comment briefly on why two other common tensor network methods appear unsuitable for a full quantum dynamics simulation of the current problem (for a more detailed comparison see~\onlinecite{comparative}).
We do not discuss \gls{mctdh} here since it was extensively explained in previous works~\cite{MEYER199073,Manthe2008,Manthe2009,2016LZG,Thoss2019isf,doi:10.1063/1.463007}.
\par
The \gls{ps} \gls{dmrg} unfolds the local Hilbert spaces of the phononic part of the system into auxiliary sites and encodes the phonon occupations into a binary representation~\cite{PhysRevB.57.6376,Zhang1999}.
Thereby, the computational complexity originating from the large local Hilbert space dimensions is reduced drastically at the cost of introducing long\hyp ranged couplings.
The \gls{ps} \gls{dmrg} has been applied successfully to study the out\hyp of\hyp equilibrium\hyp dynamics of Holstein models~\cite{Zhang1999}.
However, in these systems, the electron\hyp phonon couplings are strictly local.
For the tetracene, the situation is quite different as the diabatic states couple highly non\hyp locally to the vibrational degrees of freedom.
The arising long\hyp range couplings span the whole system such that combined with the binary encoding of the phononic Hilbert spaces, the calculations with $n_\mathrm{ph} = 63$ would require a total system size of 461 orbitals, which are coupled over the whole range of lattice sites.
It has been demonstrated~\cite{comparative} that already for groundstate calculations, the convergence of \gls{dmrg} algorithms can be quite challenging for such a large number of lattice sites in the \gls{ps} representation.
In an out\hyp of\hyp equilibrium setup, we expect these problems to become even more severe.
\par
The \gls{lbo} is another very successful method based on rotating the local phononic Hilbert space representations into an optimized basis, allowing for a truncation of the required large local dimensions~\cite{PhysRevLett.80.2661,Brockt_2015}.
This effective compression generates a significant speedup in the tensor contractions and has been applied to study the dynamics of coupled electron\hyp phonon systems following a global quench~\cite{Brockt_2015,Dorfner_2015,jansen2021chargedensitywave}.
However, in contrast to groundstate calculations, the optimization of the local basis has to be performed at each time step during the time evolution to maintain the optimal description of the vibrational degrees of freedom~\cite{Stolpp2020}.
As a consequence, only parts of the effective Hamiltonian can be represented in the optimized basis, the remainder still requires the full local Hilbert space dimension $d = n_\mathrm{ph}+1$ of the original phononic Hilbert space.
To be more precise, the most expensive numerical operation is the application of the effective Hamiltonian, which scales as $d^3$ in the local dimension.
Within the \gls{lbo} scheme, this translates into a scaling $d^2 d_{\mathrm{eff}}$, where $d_\mathrm{eff}$ is the dimension of the optimized local basis representation.
Thus, the numerical costs are reduced at the most by a factor of $d_\mathrm{eff} / d$.
Comparing this estimation for the speedup of the \gls{lbo} method with the speedup presented in \cref{fig:runtimes}, we find a significantly larger numerical efficiency exploiting the \gls{pp} mapping.
\par
Very recently~\cite{Schroeder2019} a \gls{ttns} representation has been suggested to capture the non\hyp local interactions between molecular electronic states and vibrational modes.
As we have shown in \cref{sec:numerics:stability}, a large number of excitations per vibrational mode inevitably need to be taken into account.
Exploring a \gls{ttns} representation, the idea is to reduce the numerical costs by grouping strongly correlated regions of the wave function, reducing the required bond dimension $m$.
Assuming a two\hyp site \gls{tdvp} update (of a rank--$3$ tensor), the numerically most costly operations scale as $\LandauO{m^3d^2w^2}$ or $\LandauO{m^2 d^3 w^3}$, depending on the magnitude of the local dimension $d$ and the bond dimension of the Hamiltonian representation $w$.
Using the standard tensor\hyp network representations, we found values of $w\sim\LandauO{100}$ when constructing the Hamiltonian~\cite{Hubig_2017}, i.e., a bond dimension $m\sim \LandauO{100}$ is an upper limit to maintain the numerical feasibility of the time evolution, which was also used in~\onlinecite{Schroeder2019}.
However, in a tensor\hyp network representation, the value of $m$ is directly linked to the amount of correlations that can be captured by the corresponding bond.
To resolve this conflict, entanglement renormalization by means of machine learning based optimization strategies has been applied to deduce a grouping of vibrational modes, allowing the required reduction in $m$.
In contrast, exploiting the artificial $U(1)$ symmetry introduced by the \gls{pp} mapping, we find $w\sim\LandauO{10}$, enabling us to use $m=6144$ without severe runtime losses while still working with a \gls{mps} representation, which is easily adopted to various setups.
Based on the previous discussion, we believe that combining the \gls{pp} mapping with advanced entanglement renormalization could provide a further big leap.
\section{Results}\label{sec:results}
\begin{figure}
  \centering
   \subfloat[\textwidth][\label{fig:absorption}] {
     \iflocal \input{figures/absorp_spec} \else
     \includegraphics[width=0.42\textwidth]{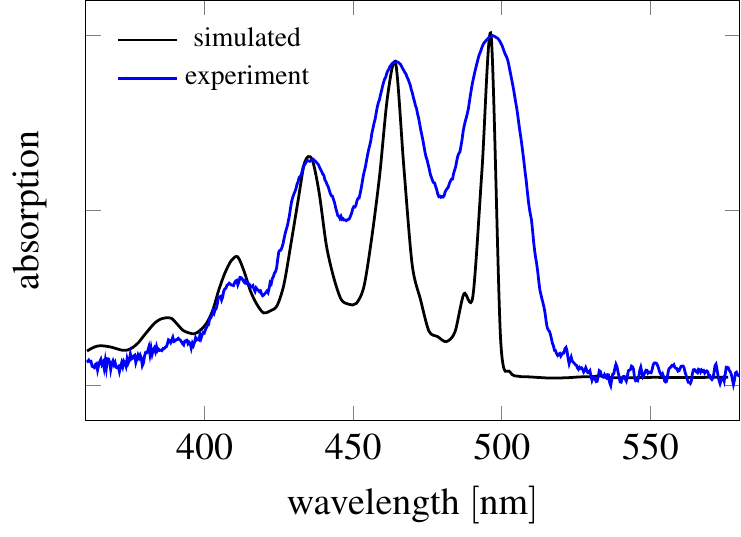} \fi
   }
   \subfloat[\textwidth][\label{fig:elec_population}] {
     \iflocal \input{figures/bright_elec_occ} \else
     \includegraphics[width=0.46\textwidth]{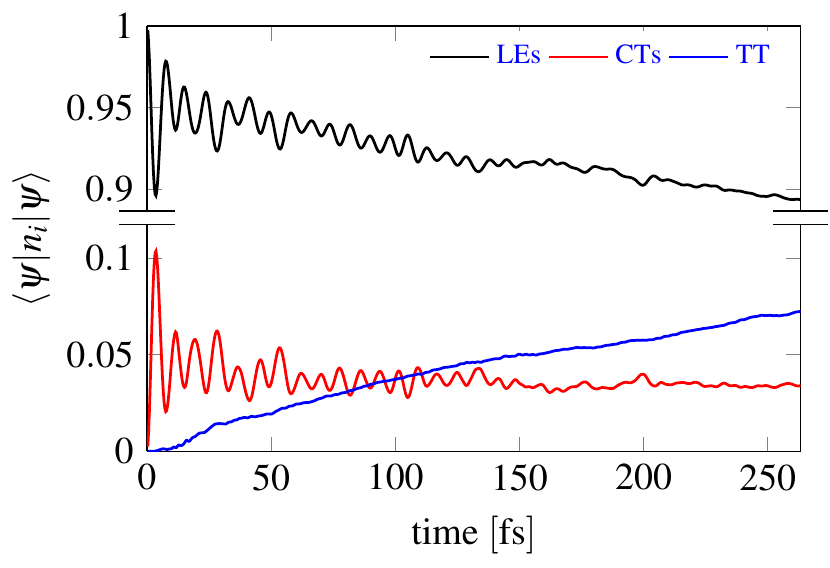} \fi
   }\\
   \subfloat[\textwidth][\label{fig:partial_energies}] {
      \iflocal \input{figures/energy} \else
      \includegraphics[width=0.46\textwidth]{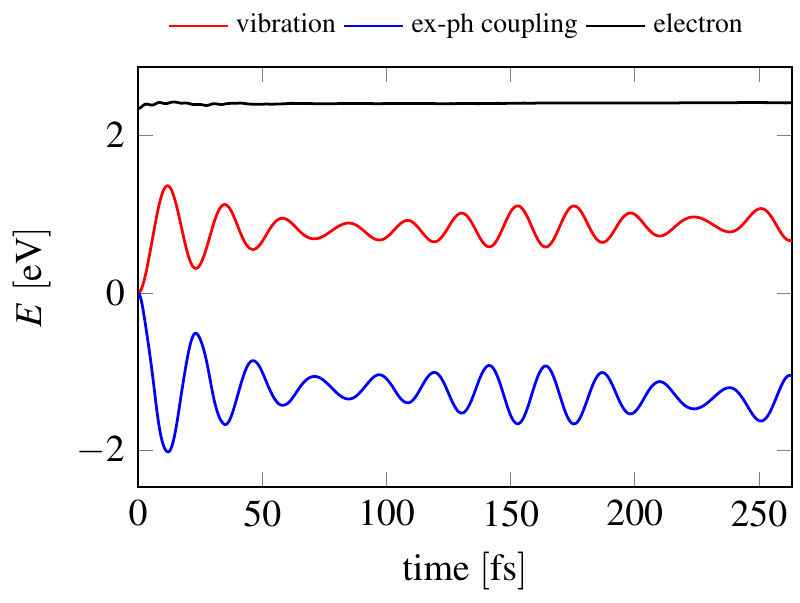} \fi
   }   
   \subfloat[\textwidth][\label{fig:partial_energies_ft}] {
     \iflocal \input{figures/energy_ft} \else
     \includegraphics[width=0.5\textwidth]{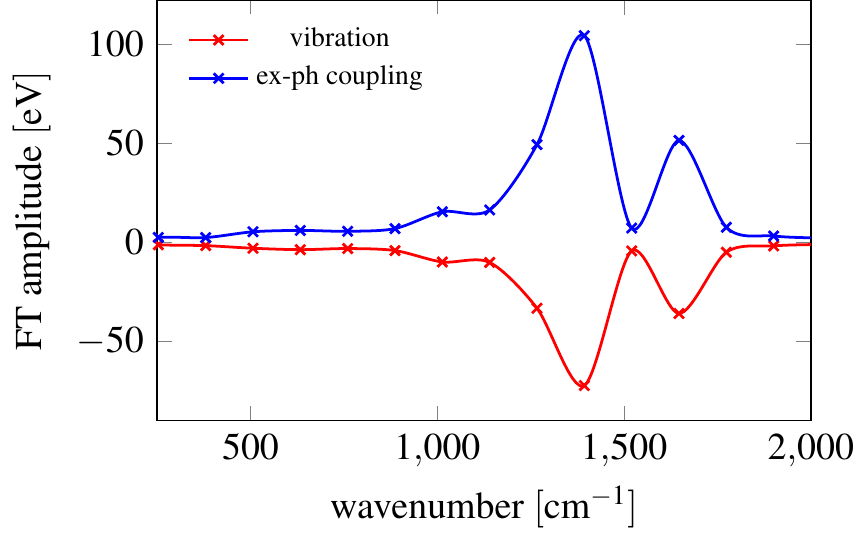} \fi
   }
  \caption{
            \label{fig:dynamics}
            (\protect\ref{fig:absorption}) Simulated absorption spectra compared with experiments. 
            (\protect\ref{fig:elec_population}) Population dynamics of different electronic states. The maximum \gls{tt} population is $\SI{7.2}{\percent}$.
            (\protect\ref{fig:partial_energies}) The expectation value of different Hamiltonian components.
            (\protect\ref{fig:partial_energies_ft}) Fourier transform of (\protect\ref{fig:partial_energies}).
          }
\end{figure}
\subsection{Absorption spectra and energy transfer}
In order to verify the applicability of our model and methods, we evaluate the absorption spectrum, \cite{Mukamel1995PrinciplesON}
\begin{align}
   \label{eq:spectral_intensity}
   I(\omega) &\propto \Re\prettyintegral{\infty}{0}{t}{C(t)\exp\left[\imagunit \omega t -t/\tau\right]}\;\, .
\end{align}
Here, $C(t)$ is the auto-correlation function,
\begin{align}
  C(t) &= \bra{\chi(0)} \bra{\varphi(0)} \enumber^{-\imagunit t H} \ket{\varphi(0)} \ket{\chi(0)}\; \text{,}
\end{align}
and $\tau$ is an appropriately short effective relaxation time.
$\ket{\varphi(0)}$ is the initial excitonic state, and $\ket{\chi(0)}$ is the vacuum phononic state.
Our system was prepared in the bright initial state configuration $\ket{\varphi(0)} = (\ket{\text{LE}_1} + \ket{\text{LE}_2})/\sqrt{2}$.
An analogous analysis for the localized and dark initial state can be found in~\suppref{sec:initialstate}.
\par
The resulting calculation is displayed in \protect\cref{fig:absorption}. In order to resemble the relative magnitude of the 0\hyp 0 and the 0\hyp 1 peaks of the experimental spectrum\cite{wang_isf_tetoli}, the relaxation time is chosen to be $\tau=\SI{230}{\femto\second}$.
We find the theoretical data capturing the main experimental features for the vibrationally resolved absorption spectrum.
In particular, the position and relative magnitudes of the several dominating vibronic peaks are in excellent agreement.
Furthermore, small wavelength features, e.g., a shoulder and a small peak (around $\sim\SI{380}{\nano\meter}$ and $\sim\SI{420}{\nano\meter}$, respectively), are found to be correct.
However, their absorption probability is slightly overestimated.
We attribute the experimentally observed increased signal widths to the neglected phononic modes, which generate additional scattering processes between diabatic states and vibronic excitations.
\par
In \protect\cref{fig:elec_population} we see the dynamics of the populations of the diabatic states.
Note that \gls{tt} has a small delay time vs \gls{ct}.
This indicates that the indirect mechanism dominates the process.
Both the populations of \gls{le} and \gls{ct} show oscillations whose frequency is about $\SI{0.12}{\per\femto\second}$.
We find this value corresponding to the energy gap between two eigenstates of the pure excitonic Hamiltonian.
These two eigenstates are accounting for over $\SI{96}{\percent}$ of \gls{le} and \gls{ct} when expanding the eigenbasis in terms of the components of the excitonic basis, i.e., they carry the main weight.
This demonstrates how far the driving force from excitonic coupling is important for \gls{isf} in this system on short transient time scales, in agreement with our findings from~\cref{subsec:solvent}.
Note that the coherent oscillations disappear in the scope of the dynamics.
This implies that the vibrational system is chosen large enough to achieve relaxation into an incoherent excitonic dynamic, contrasting previous works~\cite{2016LZG,2018JPCLPent}.
\par
Having access to the full wave function, we are able to study the energy transfer between the excitonic and phononic systems.
For that purpose, we evaluate the partial energies of the exciton and phonon system, as well as the coupling energy by separating~\cref{eq:Hamiltonian}.
The obtained time dependencies are displayed in~\cref{fig:partial_energies}.
Most importantly, we find only few dominating oscillations in the coupling and vibrational energies at finite values, after an initial transient regime.
As shown in \cref{fig:partial_energies_ft}, a Fourier transformation reveals that there are two frequencies dominating the energy conversion between the exciton\hyp phonon coupling and the phononic subsystem.
They are located at around $\SI{1420}{\per\centi\meter}$ and $\SI{1620}{\per\centi\meter}$, respectively.
We identify these signals with vibrational modes as follows:
\begin{align*}
	\SI{1420}{\per\centi\meter} &: &\text{modes no. 184, 185 and 186} \; \text{,} \\
	\SI{1620}{\per\centi\meter} &: &\text{modes no. 209, 210 and 211} \; \text{.}
\end{align*}
Note that all of these frequencies correspond to the strong peaks in the spectral densities in~\cref{fig:spectral_density_lele,fig:spectral_density_tttt}.
The observation that modes 184-186 are most relevant for the energy transfer can be related to the nature of these vibrational modes, i.e., they correspond to collective vibrations of the whole molecule's backbone (see~\suppref{sec:vibmodes}).
We note that previously~\cite{2016LZG,Bakulin2016Nature,schnedermann2019} studies of the vibrational contributions to the excitonic dynamics found a similar relation for vibrational modes in the regime of $\SI{1420}{\per\centi\meter}$.
Here, the importance of only few vibrational modes was determined either by successively eliminating oscillation frequency windows~\cite{2016LZG}, or by combined experimental and theoretical analyses on a related molecule~\cite{Bakulin2016Nature,schnedermann2019}.
\subsection{Solvent and triplet energy dependence}\label{subsec:solvent}
In a subsequent stage, we investigated the dependency of the triplet production on the energies of \glspl{ct} and \gls{tt}.
Importantly, the \glspl{ct} on\hyp site energies can be tuned in the experiment by using various solvents with different polarities~\cite{2020JACSsolvent}.
For instance, Alvertis et al.~\cite{doi:10.1021/jacs.9b05561} recently revealed the significant energy splitting of \gls{ct} states of the \gls{dt-mes} in polar solvents and its vital role in the switching between coherent and incoherent iSF.
However, because the inter\hyp chromophore couplings are much smaller in our covalently linked tetracene dimer system than in \gls{dt-mes}, due to an additional phenylene group between two tetracene units, the local electric field by polar solvents will not result in significant differences between the two \gls{ct} states in our case.
Therefore, the energy splitting between the two \gls{ct} states is neglected in this work.
\newline
\begin{figure}[!ht]
  \centering
  \subfloat[\label{fig:ct_tt_mat_plot}] {
    \iflocal \input{figures/solvent_tt_occ} \else
    \includegraphics{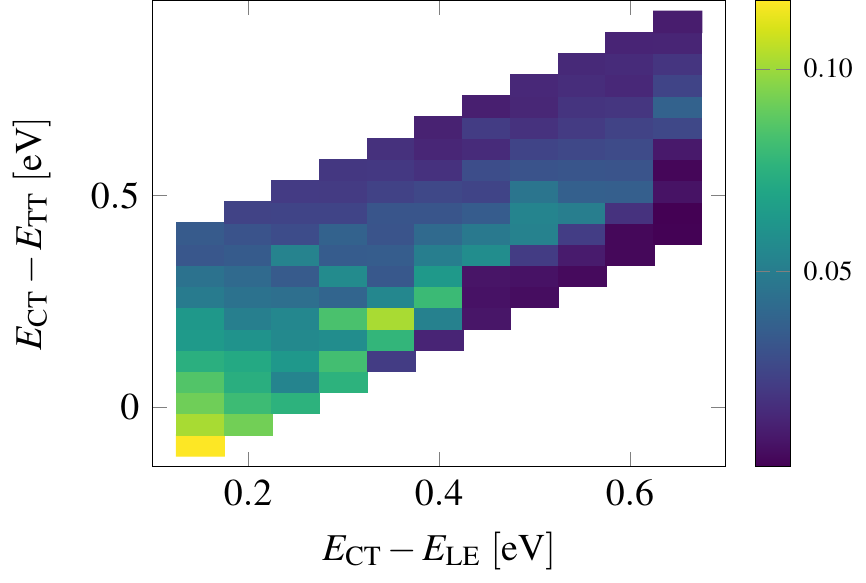} \fi
  }
  \subfloat[\label{fig:ct_tt_plot}] {
    \iflocal \input{figures/selected_solvent_tt_occ} \else
    \includegraphics{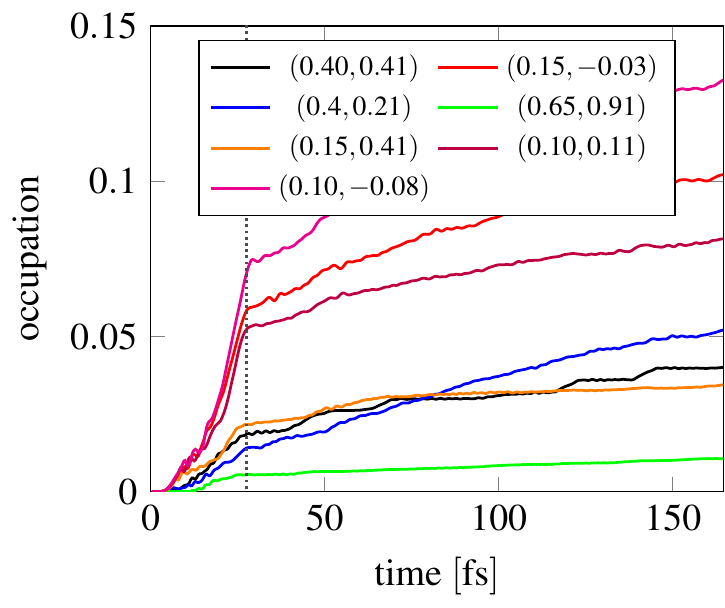} \fi
  }
  \caption{\label{fig:solvent_tt_occ}\protect\cref{fig:ct_tt_mat_plot} shows the maximum \gls{tt} population for different energies of the exciton states. Notice, that for each energy pair the maximum value is chosen, i.e. the occupations do not belong to the same point in time. \protect\cref{fig:ct_tt_plot} gives the \gls{tt} population with respect to the excitonic energies $(\Delta E(\mathrm{CT} - \mathrm{LE}), \Delta E(\mathrm{CT} - \mathrm{TT}))$ in $\si{\electronvolt}$.}
\end{figure}
As shown in~\cref{fig:solvent_tt_occ} we scanned a whole parameter region shifting the energies.
We measure the offsets of the \glspl{ct} and \gls{tt} from the values given in~\cref{table:electronic_coupl} by the gaps $\Delta E(\mathrm{CT}-\mathrm{LE})$ and $\Delta E(\mathrm{CT}-\mathrm{TT})$, respectively.
\par
Most prominently, we find that lowering the energy of the \gls{ct} states, i.e., increasing the solvent polarity, yields a larger \gls{tt} population rate at the early stage.
Moreover, the gained triplet population can be elevated by increasing the energy of the \gls{tt} state.
Combining these observations, we conclude that one dominating process for the \gls{tt} yield is the relaxation of \gls{ct} populations into the energetically closest diabatic states, suggesting the indirect \gls{isf} mechanism (\gls{le}$\rightarrow$\gls{ct}$\rightarrow$\gls{tt}).
This picture is consistent with the observation that the \gls{tt} production rate is strongly suppressed in the lower right corner of~\cref{fig:ct_tt_mat_plot}.
In addition, we note that the triplet yield obtained from the system parameters (\cref{table:electronic_coupl}) matches the experimental findings~\cite{wang_isf_tetoli}, i.e., we find a relatively small value of $\SI{14}{\percent}$.
\par
Previous studies~\cite{Chan1541,doi:10.1146/annurev-physchem-040214-121235,2016LZG,2018JPCLPent} found a finite time scale for the coherent dynamics.
In order to determine this electronic coherence time, in~\cref{fig:ct_tt_plot}, we show the time\hyp evolution of the \gls{tt} population for selected limiting cases.
We find a consistent change in the dynamics at $t_0 \sim\SI{35}{\femto\second}$ where the slope of the \gls{tt} production rate changes independently of the specific choice of the on\hyp site potentials.
As discussed in~\suppref{sec:rdmanalysis}, this can be addressed to a change in the nature of the excitonic dynamics.
At times $t<t_0$, all excitonic states realize a coherently driven dynamics.
However, at times $t>t_0$, this mechanism changes into a classically driven relaxation of mixed states of the \glspl{le}, \glspl{ct}, and \gls{tt} subsystems.
Previously, the coherent increase in the \gls{tt} population has been interpreted in terms of phonon\hyp assisted hoppings employing Fermi's golden rule~\cite{2016LZG}.
However, we also found that the electronic coherence time roughly coincides with the oscillation time of the dominating vibrational mode nos. 184 and 185.
In fact, their wavenumber of $\SI{1420}{\per\centi\meter}$ roughly translates to a period of $\SI{25}{\femto\second}$.
Noting also that $\sim 2/3$, i.e., the dominating part of the \gls{tt} population is generated during the coherent dynamics, this could create a competition between phonon assisted hoppings and the time scale in which these processes are possible.
In \suppref{sec:rdmanalysis}, we investigated the renormalization of the excitonic couplings after rewriting the Hamiltonian using a Lang\hyp Firsov transformation~\cite{lang1963kinetic}.
Importantly, we find that the renormalization of the couplings between the \gls{tt} state and the remaining excitonic system changes from an enhancement to a suppression after $t_0 = \SI[separate-uncertainty=true]{34.725(165)}{\femto\second}$.
It is a remarkable observation that the description of the phonon\hyp assisted hopping in terms of polaron dynamics yields a perfectly matching prediction of the electronic coherence time.
In this quasi\hyp particle picture, the coherent phonon\hyp assisted hopping processes are blocked by the momentum transfer into the vibrational system.
In fact, this momentum transfer creates a cloud of vibrations accompanying the excitonic states~\cite{Brockt_2015}, and thereby increases the effective exciton mass.
Our analysis further shows that within the coherent dynamics, the dominating hopping is between \gls{ct} and \gls{tt} states, pointing out the importance of the indirect hopping mechanism.
\begin{figure}
  \centering
  \includegraphics{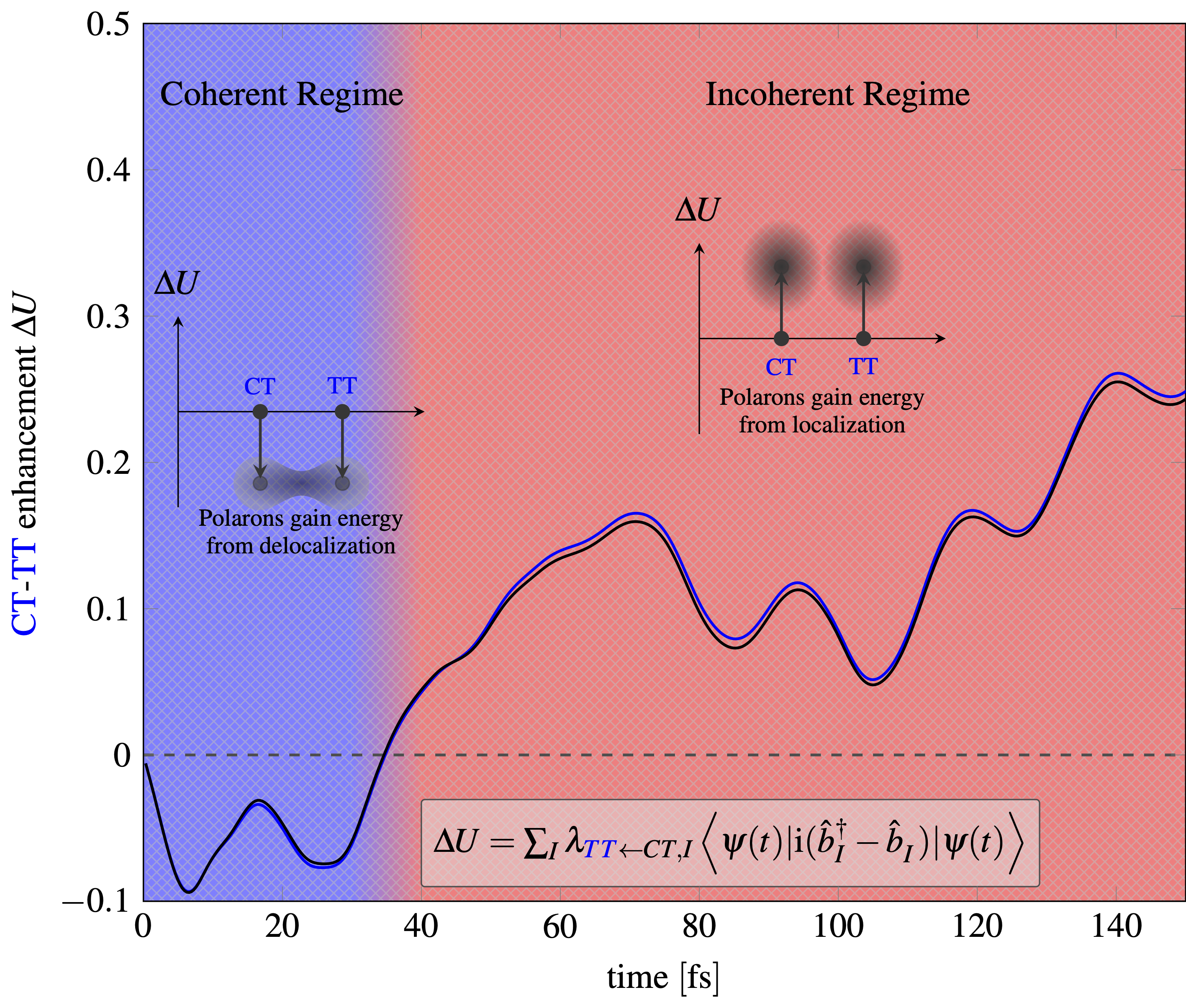}
  \caption{\label{fig:polaron-sketch}\gls{ct}\hyp \gls{tt} hopping enhancement with the schematic sketch of quasi\hyp particle formation.
    The coefficients $\lambda_{ij,I} = g_{ij,I}/\omega_I$ define the ration between the exciton\hyp phonon coupling and the vibration frequencies.
  }
\end{figure}
A schematic representation of the different regimes and the corresponding type of quasi\hyp particles governing the dynamics is shown in~\cref{fig:polaron-sketch}.
\section{Conclusion and Outlook}
We studied the \gls{tt} production rate via \gls{isf} for a covalently linked tetracene dimer using a large\hyp scale full quantum\hyp mechanical treatment of the relaxation dynamics after photo\hyp excitation.
Our study extends the theoretical analysis conducted so far~\cite{2016LZG} by treating couplings between the diabatic states and the vibrational modes obtained from first principles.
Incorporating a large number of molecular vibrational modes, we achieve a realistic description of the full molecule's dynamics, following photo\hyp excitation.
We obtain excellent agreement with experimentally observed absorption spectra and identify only a small number of vibrational modes, dominating the energy conversion between the different subsystems.
Analyzing the renormalization of the exciton couplings in terms of exciton\hyp phonon quasi\hyp particles, we can trace back the origin of the phonon\hyp assisted coherent \gls{tt} generation to an enhanced delocalization of the created quasi\hyp particles, which after some time transform into heavy, localized ones.
A systematic variation of molecular parameters that can be related to the solvent polarity allows us to identify those regions in parameter space exhibiting the largest triplet yield.
\par
As it is of special interest to raise the \gls{tt} population to produce a high charge carrier yield, our findings can be employed to identify promising manipulations of the molecular structure.
Here, we suggest two pathways.
First of all, decreasing the energy gaps between \gls{ct} and \gls{le} or \gls{tt} is crucial for producing the higher \gls{tt} yield.
Our results indicate that both modifications of \gls{ct} and \gls{tt} energies mainly affect the overall scale of the triplet yield.
Here, during the initial coherent dynamics, a quick increase in the \gls{tt} is observed which transforms to a classically driven relaxation, happening on time scales $t\gtrsim t_0 = \SI{35}{\femto\second}$.
We note that the observed time scale for coherent dynamics is in agreement with previous studies, both theoretical and experimental~\cite{wang_isf_tetoli,2015PRLJAP,2016LZG,2018JPCLPent,2013AccountsSF,2019TGHJCP,2017JHLPCL,doi:10.1021/acs.jctc.9b00122,2015PRLJAP,monahannature}.
However, this time scale is very close to the oscillation period $\sim\SI{25}{\femto\second}$ of the vibrational mode nos. 184, 185, and 186 dominating the energy transfer between the excitonic and vibrational systems.
Therefore, our findings suggest a second route that, though experimentally more challenging, would be an increase in the electronic coherence time $t_0$.
Noting that the dominating fraction of the \gls{tt} yield is generated from the initial, coherent dynamics, a combination of larger coherent times, i.e., a reduction in the frequency of the oscillations of the overall molecule's backbone with an increased solvent polarity could, generate significantly larger \gls{tt} occupations.
\par
There are still many aspects to be further considered for future simulations of \gls{sf} dynamics.
First of all, we might take into account more chromophore units.
It was suggested recently that the spatial separation of the correlated triplets might be a crucial point in the harvest of a large yield~\cite{wang_isf_tetoli}.
Since the presented numerical framework can adaptively reduce the number of vibrational modes, it might be a promising tool to explore the dynamics of larger molecules, too.
Additionally, finite temperature effects, which may become relevant for larger molecules~\cite{wang_isf_tetoli}, can be incorporated by a thermofield approach, which is a standard technique for tensor network methods.
From a methodical point of view, a direct comparison between established time\hyp evolution schemes such as multi\hyp layer \gls{mctdh}~\cite{Manthe2008,Manthe2009} and the presented tensor network representation would be desirable.
In particular, a faithful investigation of these different approaches could yield insights which method should be used to optimally describe \gls{sf} dynamics.
Furthermore, we believe that a combination of our representation of the vibrational degrees of freedom with \gls{ttns} and entanglement renormalization techniques~\cite{Schroeder2019} could open the path to study the full quantum dynamics of generic, large organic semi\hyp conductors.
\section*{Supplementary Material}
See Supplementary Material for details on the derivation of the model, as well as details on the numerics during the simulation.
Furthermore it contains a detailed analysis of the electronic \gls{rdm} in combination with the derivation of the effective renormalized couplings.
\begin{acknowledgments}
We thank Dr. Xiaoyu Xie, Dr. Thomas K\"ohler and Professor Uwe Manthe for helpful discussions.
We acknowledge financial support by the Deutsche Forschungsgemeinschaft (DFG, German Research Foundation) under Germany's Excellence Strategy-426 EXC-2111-390814868 (U.S.) and the National Natural Science Foundation of China (Grant No. 22073045) (Y.X., X.Y. and H.M.).
Furthermore, we acknowledge support from the Munich Center for Quantum Science and Technology (S.M., M.G., U.S, and S.P.).
All calculations made use of the SyTen toolkit~\cite{hubig:_syten_toolk}, and some comparison benchmarks were performed using the SymMPS toolkit~\cite{symmps}.
\end{acknowledgments}
\section*{Data Availability Statement}
The data that support the findings of this study are available from the corresponding author upon reasonable request.
\section*{Author contributions}
S.M. developed the idea, wrote the code and performed the calculations.
Y.X. X.Y. and H.M. guided with chemical intuition to the chosen molecule.
Y.X. and X.Y. did electronic structure calculations and the computation of the vibrational modes.
S.M., Y.X. and S.P. interpreted the results and wrote the paper together.
S.P. supervised the project and is one of the original authors of the main method.
H.M., M.G. and U.S. discussed the results and helped interpret the data.
\section*{Additional information}
%
%
\subsection*{Competing interests}
The authors declare no competing interests.
\appendix
\section{Electronic Structure Calculation}\label{sec:elecstruct}
We employ \gls{casscf} with a $6-31G(d)$ basis set in order to calculate all the excited states involved in our \gls{isf} model due to the multi\hyp excitation nature of \gls{tt}.
The four orbitals near \gls{homo} and \gls{lumo} are delocalised.
Therefore we use the Pipek\hyp Mezey method \cite{PMLocalisation} on these orbitals to generate four localised ones (from \gls{homo}-1 to \gls{lumo}+1).
Each of them is confined in single tetracene chromophore.
We construct all the diabatic states from the four localised orbitals and calculate all the excitonic Hamiltonian elements $V_{ij}$ using \gls{casci}.
All calculations are performed using the OpenMolcas package~\cite{openmolcas}.
\par
Here we only take into account the linear exciton-phonon coupling terms.
We distort the whole molecule (at equilibrium geometry) along selected vibrational modes' displacement for $\pm\SI{0.01}{\angstrom}$.
Then we calculate the excitonic Hamiltonian with the same procedure as mentioned above.
With the Hamiltonian at three distinct points in space ($\pm\SI{0.01}{\angstrom}$,$\SI{0}{\angstrom}$) , we can perform linear fitting of each elements $V_{ij}$ and calculate the gradient.
Note that the coupling $g_{ij,I}$  in~\cref{eq:ex-ph} is connected to dimensionless displacement $Q_I$.
One has to multiply the gradient by a factor of $\sqrt{\hbar/\left(m_I\omega_I\right)}$. 
\begin{align}
   \label{eq:ex-ph}
   g_{ij,I} &= \nabla_{Q_{I}}V_{ij}(\mathbf{Q})
\end{align}
\section{Vibrational modes}\label{sec:vibmodes}
\begin{figure}
  \centering
   \subfloat[\textwidth][\label{fig:mode1}] {
     \includegraphics[width=0.2\textwidth]{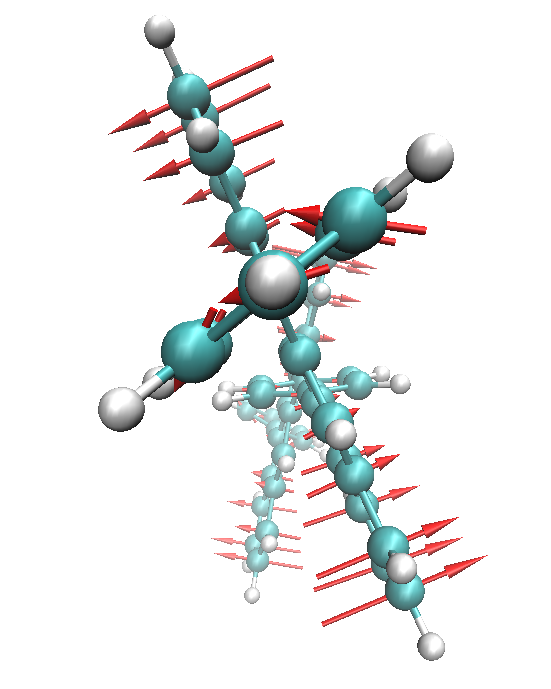}
   }
   \subfloat[\textwidth][\label{fig:mode45}] {
     \includegraphics[width=0.39\textwidth]{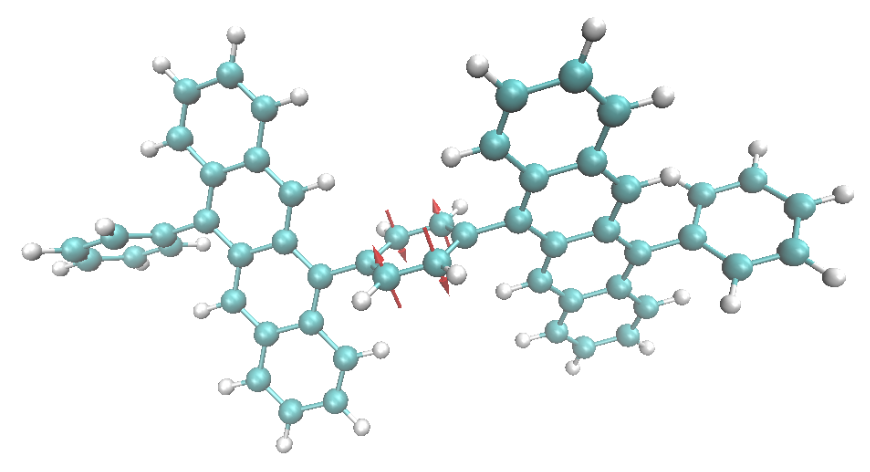}
   }\\
   \subfloat[\textwidth][\label{fig:mode167}] {
     \includegraphics[width=0.37\textwidth]{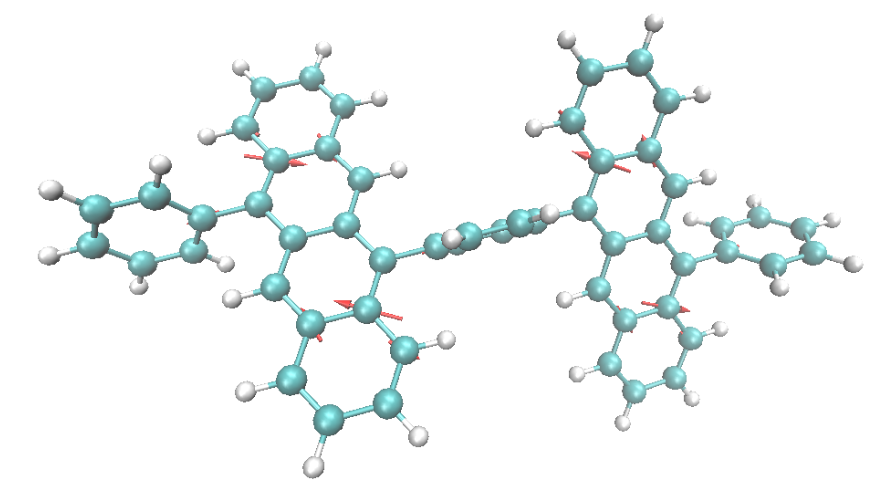}
   }
   \subfloat[\textwidth][\label{fig:mode168}] {
     \includegraphics[width=0.39\textwidth]{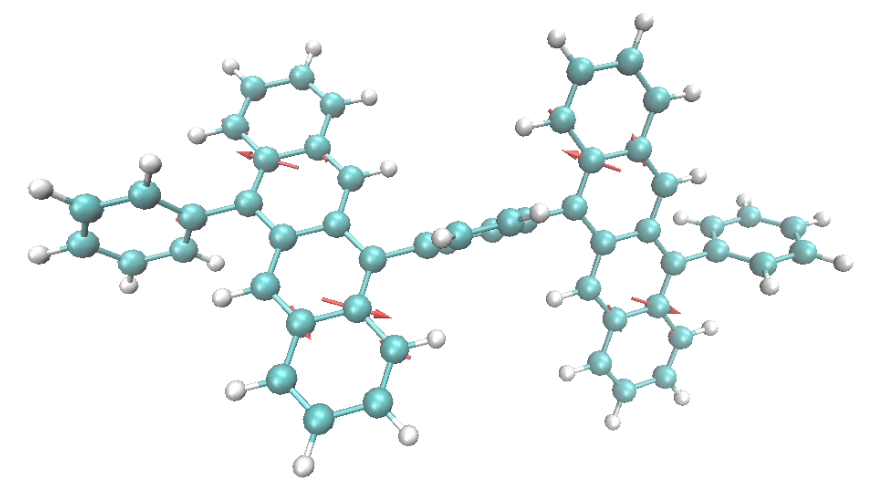}
   }\\
    \subfloat[\textwidth][\label{fig:mode184}] {
     \includegraphics[width=0.37\textwidth]{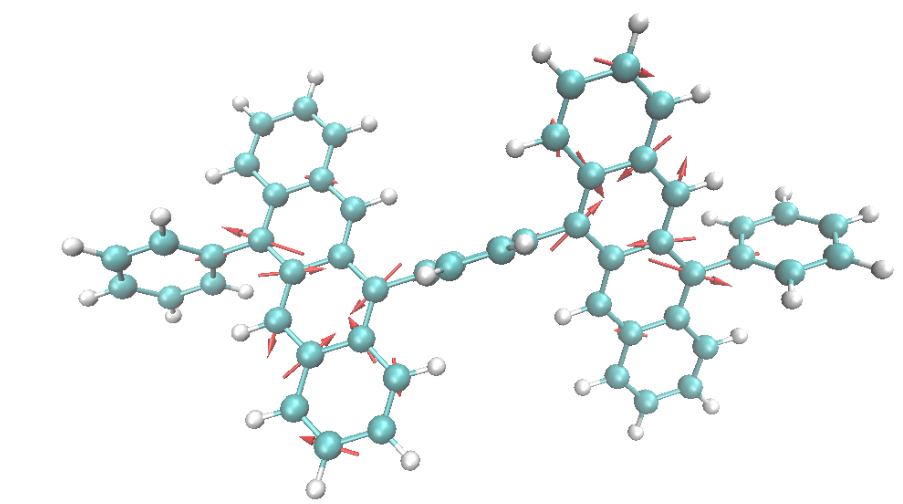}
   }
   \subfloat[\textwidth][\label{fig:mode185}] {
     \includegraphics[width=0.39\textwidth]{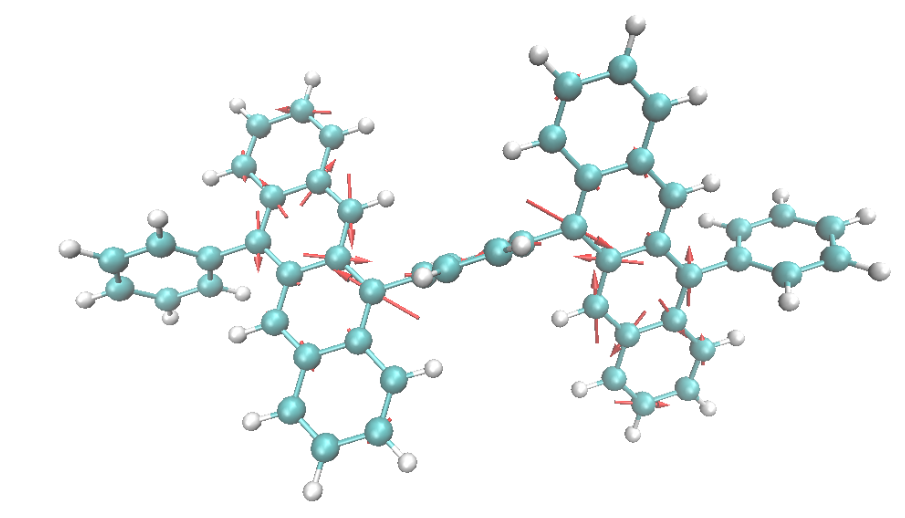}
   }\\
   \subfloat[\textwidth][\label{fig:mode210}] {
     \includegraphics[width=0.37\textwidth]{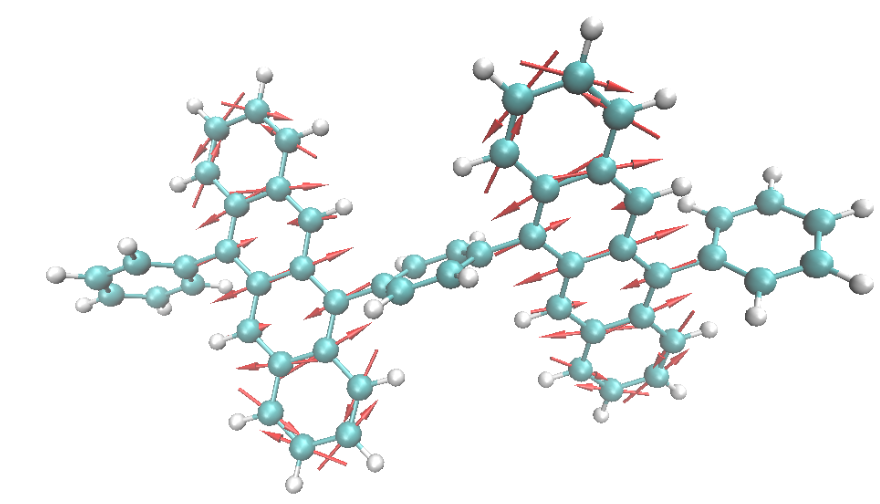}
   }
   \subfloat[\textwidth][\label{fig:mode211}] {
     \includegraphics[width=0.39\textwidth]{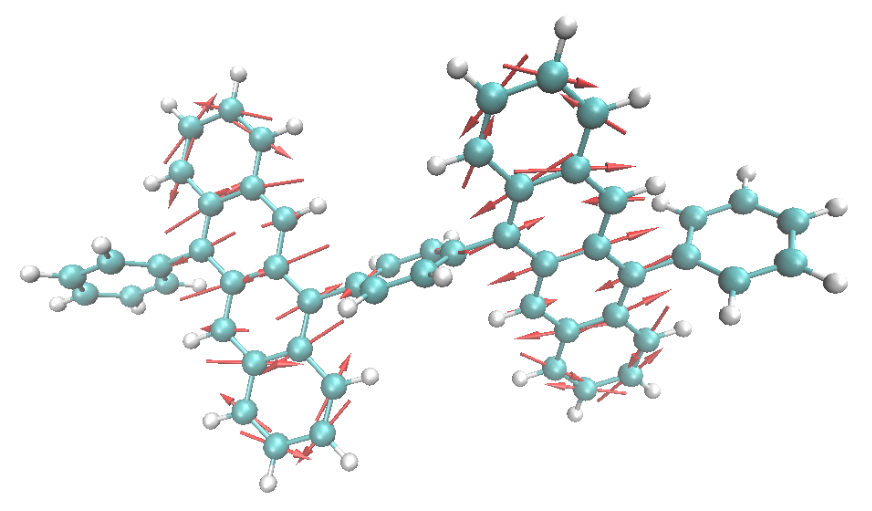}
   }
  \caption{\label{fig:vibrational_modes} Illustration of displacement vector of some essential vibrational modes:
            (\protect\ref{fig:mode1}) mode no. 1 ($\SI{6.7}{\per\centi\meter}$).
            (\protect\ref{fig:mode45}) mode no. 45 ($\SI{415.0}{\per\centi\meter}$).
            (\protect\ref{fig:mode167}) mode no. 167 ($\SI{1277.0}{\per\centi\meter}$).
            (\protect\ref{fig:mode168}) mode no. 168 ($\SI{1277.7}{\per\centi\meter}$).
            (\protect\ref{fig:mode184}) mode no. 184 ($\SI{1409.6}{\per\centi\meter}$).
            (\protect\ref{fig:mode185}) mode no. 185 ($\SI{1411.2}{\per\centi\meter}$).
            (\protect\ref{fig:mode210}) mode no. 210 ($\SI{1624.9}{\per\centi\meter}$).
            (\protect\ref{fig:mode211}) mode no. 211 ($\SI{1625.4}{\per\centi\meter}$).}
\end{figure}
\clearpage
\section{Spectral Density Data}\label{sec:specdensity}
\begin{figure}[!h]
  \centering
   \subfloat[0.45\textwidth][\label{fig:spectral_density_le1le2}] {
     \iflocal \input{figures/spectral_density_le1le2} \else
     \includegraphics[width = 0.46\textwidth]{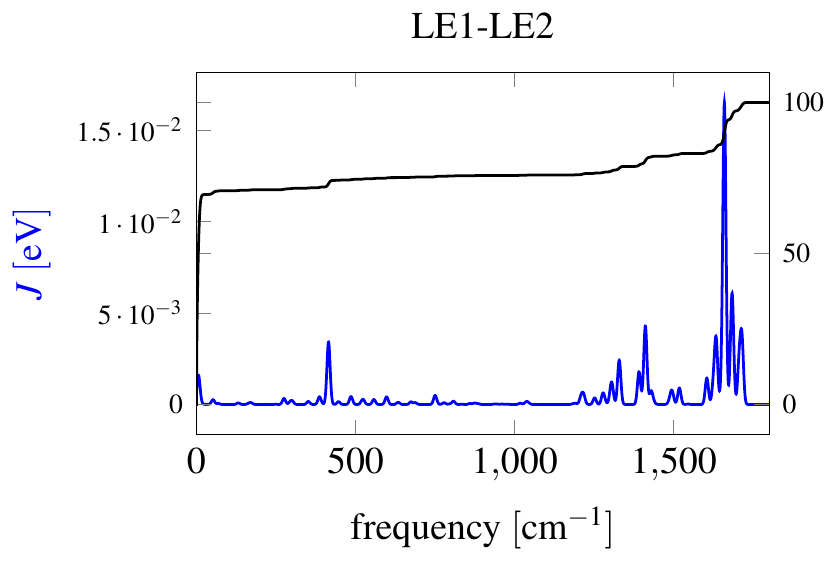} \fi
   }
   \subfloat[0.45\textwidth][\label{fig:spectral_density_le1ct2}] {
     \iflocal \input{figures/spectral_density_le1ct2} \else
     \includegraphics[width = 0.45\textwidth]{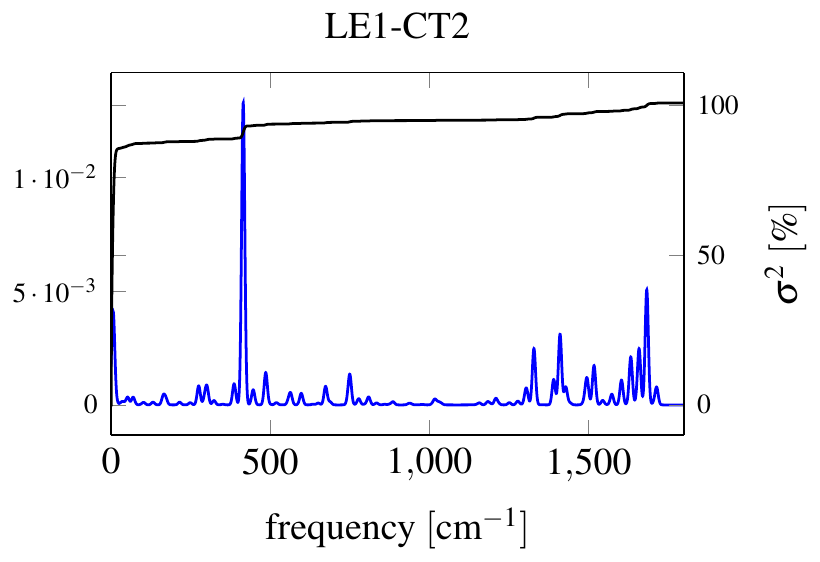} \fi
   }\\
   \subfloat[0.45\textwidth][\label{fig:spectral_density_ct1ct1}] {
     \iflocal \input{figures/spectral_density_ct1ct1} \else
     \includegraphics[width = 0.42\textwidth]{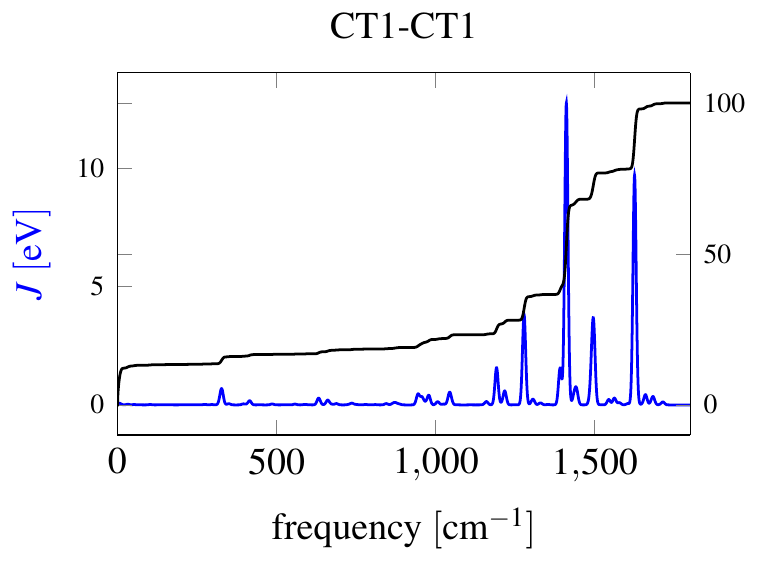} \fi
   }
   \subfloat[0.45\textwidth][\label{fig:spectral_density_ct1ct2}] {
     \iflocal \input{figures/spectral_density_ct1ct2} \else
     \includegraphics[width = 0.45\textwidth]{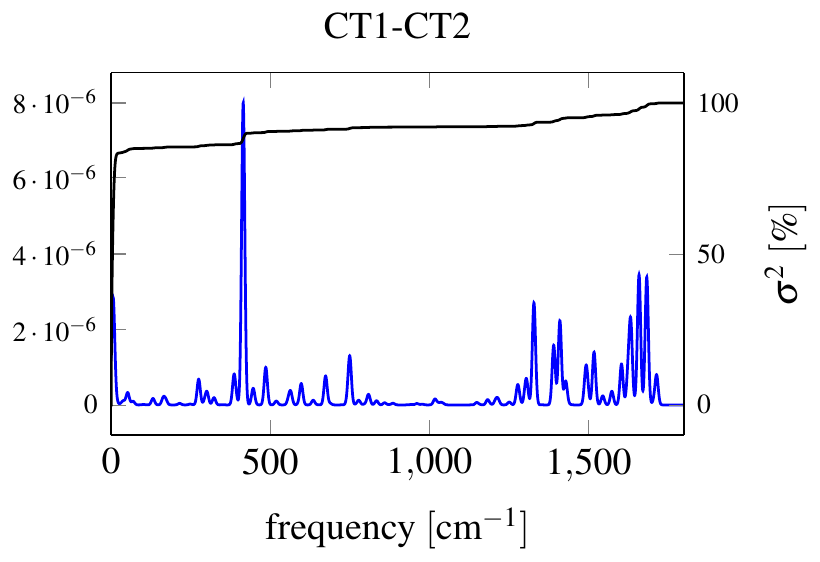} \fi
   }\\
   \subfloat[0.45\textwidth][\label{fig:spectral_density_le1tt}] {
     \iflocal \input{figures/spectral_density_le1tt} \else
     \includegraphics[width = 0.45\textwidth]{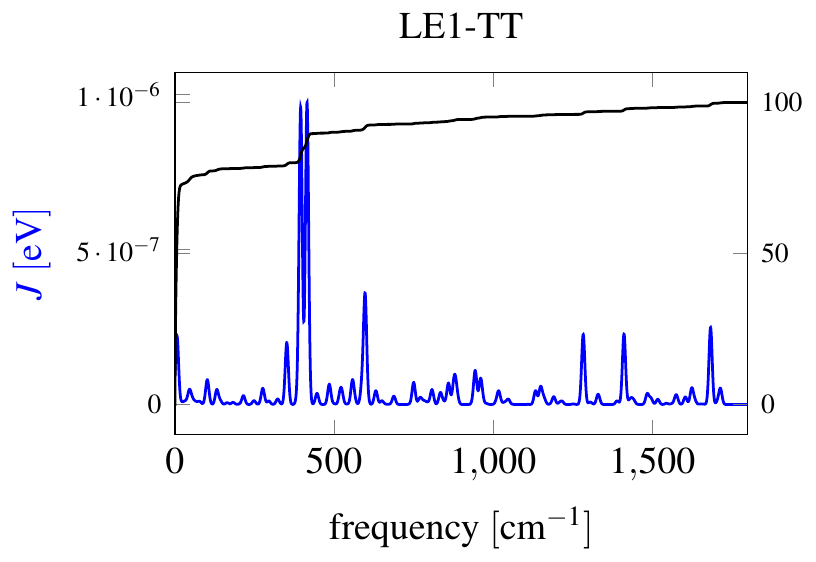} \fi
   }
   \subfloat[0.45\textwidth][\label{fig:spectral_density_le2tt}] {
     \iflocal \input{figures/spectral_density_le2tt} \else
     \includegraphics[width = 0.45\textwidth]{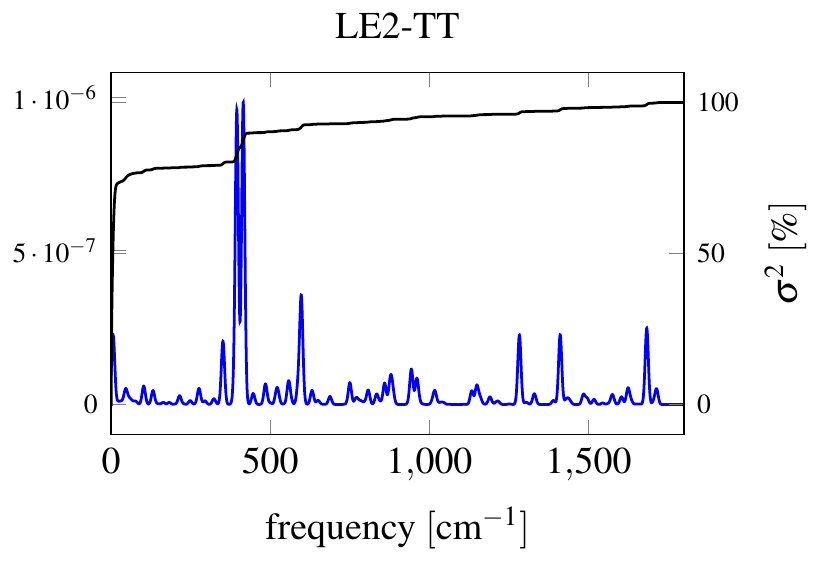} \fi
   }
  \caption{\label{supp-fig:spectral_densities}Spectral density (blue, left axis) and accumulative fluctuations (black, right axis) of selected excitonic Hamiltonian terms at room temperature.}
\end{figure}

Spectral densities and fluctuations for some excitonic Hamiltonian elements which were not included in the main text are displayed in \cref{supp-fig:spectral_densities}.
\section{Redfield search}\label{sec:redfieldsearch}
To obtain a reasonable shifted \gls{ct} energy, we use Redfield theory~\cite{REDFIELD19651,MRT2003} to simulate the \gls{sf} dynamics at the temperature of $\SI{300}{K}$.
In Redfield theory, the dynamics are described using
\begin{align}
\frac{\partial\rho_{\alpha\beta}}{\partial t} =
\imagunit(\epsilon_{\alpha}-\epsilon_{\beta})\rho_{\alpha\beta}+\sum_{\alpha\beta\gamma\kappa} R_{\alpha\beta\gamma\kappa}\rho_{\gamma\kappa} \; \text{,}
\end{align}
where $\rho$ is the density matrix of the excitonic states, and $\alpha,\beta,\gamma,\kappa$ label eigenstates of the excitonic Hamiltonian.
Furthermore the Redfield tensor $R_{\alpha\beta\gamma\kappa}$ is defined as
\begin{align}
R_{\alpha\beta\gamma\kappa} &= \Gamma^{+}_{\kappa\beta\gamma\alpha}
                                  + \Gamma^{-}_{\kappa\beta\gamma\alpha}
-\delta_{\kappa\beta}\sum_{\lambda}\Gamma^{+}_{\alpha\lambda\lambda\gamma}
-\delta_{\gamma\alpha}\sum_{\lambda}\Gamma^{-}_{\kappa\lambda\lambda\beta} \; \text{,} \\
\Gamma^{+}_{\kappa\beta\gamma\alpha} &= \prettyintegral{t}{0}{\tau}
{\exp\left\{i(\epsilon_{\gamma}-\epsilon_{\kappa})\tau\right\}}\sum_{j}C_{j}(\tau)
\braket{\alpha}{j}
\braket{j}{\beta}
\braket{\gamma}{j}
\braket{j}{\kappa} \; \text{,} \\
\Gamma^{-}_{\kappa\beta\gamma\alpha} &= \prettyintegral{t}{0}{\tau}
{\exp\left\{i(\epsilon_{\alpha}-\epsilon_{\beta})\tau\right\}}\sum_{j}C_{j}(\tau)
\braket{\alpha}{j}
\braket{j}{\beta}
\braket{\gamma}{j}
\braket{j}{\kappa} \; \text{.}
\end{align}
$\epsilon$ is the energy of the corresponding eigenstate and $C_{j}=\braket{\delta{\epsilon}_j(t)}{\delta{\epsilon}_j(0)}$ is the energy fluctuation correlation function of the state $j$.
We then tune the energy of \gls{ct} within the range of $\SI{3.0645}-\SI{3.5645}{\electronvolt}$, according to the \gls{casscf} results and the previous solvent effect investigations \cite{wang_isf_tetoli}, while fixing the energy of the other excitonic states.
In order to perform the Redfield dynamics we choose the bright excited state as was discussed in the main text.
Since $E_\mathrm{CT} = \SI{3.3645}{\electronvolt}$ has the highest \gls{tt} population we choose this as the \gls{tdvp} input. 
\begin{table}
  \centering
  \begin{tabular}{l|r}
    \hline
    \gls{ct} energy $[\si{\electronvolt}]$ & \gls{tt} population $[\si{\percent}]$
    \\
    \hline
    3.0645 &  7.72 \\
    3.1645 &  0.43 \\
    3.2645 &  13.8 \\
    3.3645 &  34.78 \\
    3.4645 &  0.04 \\
    3.5645 &  0.02 \\
    \hline
  \end{tabular}
  \caption{\label{table:redf} \gls{tt} population for different \gls{ct} energies during the Redfield search.}
\end{table}
\section{\label{numcontrolparams}Numerical control parameters}
\begin{figure}[!h]
      \centering
      \iflocal \input{figures/numerical_control_params} \else
      \includegraphics{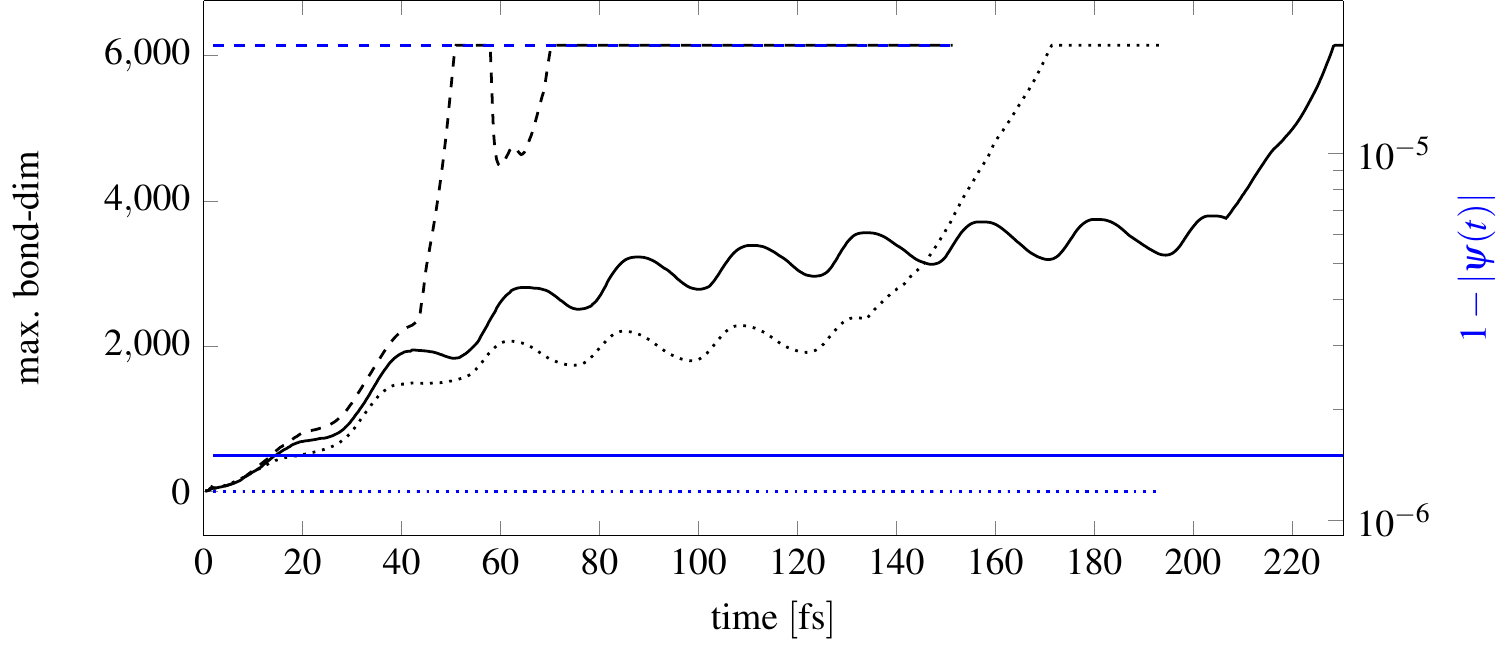} \fi
      \caption{\label{fig:numerical_control_params}Numerical control parameters plotted against time. The maximum bond dimension (black) is plotted for the localised (dotted), bright (solid) and dark (dashed) initial state. The breaking of unitarity during the \gls{tdvp} is observed through the proxy of the initial norm deviation (blue). All of these curves were obtained with $E_\mathrm{TT} = \SI{2.9493}{\electronvolt}$.}
  \end{figure}

In order to convince ourselves of the validity of the simulations we plot the maximum bond dimension and norm deviation in~\cref{fig:numerical_control_params}.
As we can see the localised and bright initial state have a rather moderate growth in the bond dimension.
Therefore they are exact, with respect to the truncated weight of $\delta = \num{1.e-8}$, until almost the end of the simulation time.
In contrast the dark initial state almost immediately skyrockets and always satisfies the upper bound.
Notice the interesting kink at which the maximum bond dimension drops here for a short time.
It is important to mention that only few of the modes actually reached the saturation of the bond dimension limit.
Typically these are the vibrational modes, whose spectrum decays very slowly, e.g. no. 184 and 185.
Even though the memory requirement depends on the exact model parameters, for those presented in~\cref{table:electronic_coupl} the final wave function (living in a formally $\sim 10^{137}$ dimensional Hilbert space) used only $\SI{1.3}{\giga\byte}$ on the disk.
This is to be compared to~\onlinecite{Schroeder2019} where the central tensor had a requirement of $\SI{57}{\tera\byte}$.
\par
\gls{tdvp} is an inherently local method which is only exact for nearest\hyp neighbour Hamiltonians or infinite bond dimension~\cite{TDVPreview}.
Otherwise a so called projection error is introduced, which causes the time evolution to be non\hyp unitary.
In order to estimate this error one can look at the deviance from the initial norm as in~\cref{fig:numerical_control_params}.
As we can recognise, the total norm difference is constant throughout the simulation.
Furthermore, this is in accordance with the truncation error.
However, it is indeed initial state dependent, i.e. the bright exciton initial state has twice the norm error of the localised one while the dark initial state actually lies one order of magnitude higher.
This is explained with regard to the bond dimension, i.e. the phononic modes needing high bond dimensions.
Since the \gls{tdvp} projector is exact for an infinite bond dimension, i.e. for an exact state~\cite{TDVPreview}, the stronger the truncation of the state, the stronger the breaking of unitarity.
\section{Initial state dependence}\label{sec:initialstate}
\begin{figure}[!h]
  \centering
  \subfloat[0.5\textwidth][\label{fig:loc_elec_population}] {
    \iflocal \input{figures/localized_elec_occ} \else
    \includegraphics[width = 0.5\textwidth]{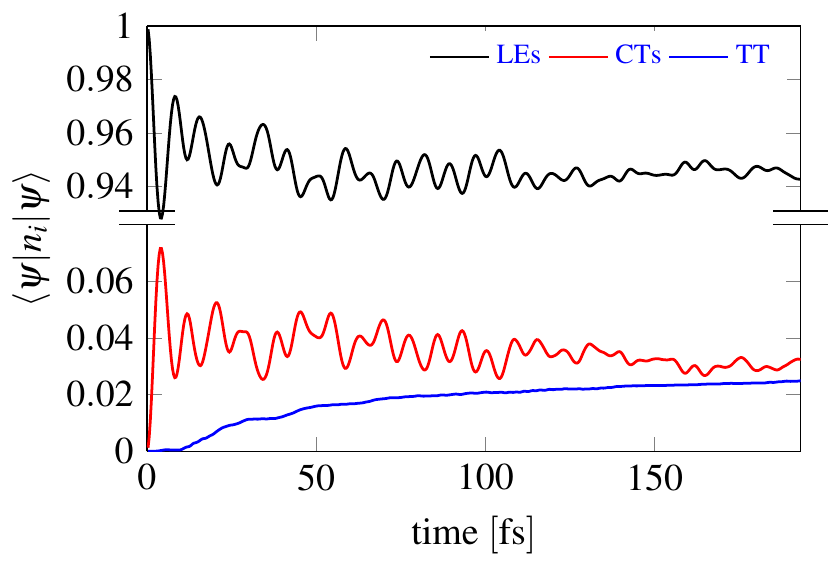} \fi
  }
  \subfloat[0.5\textwidth][\label{fig:dark_elec_population}] {
    \iflocal \input{figures/dark_elec_occ} \else
    \includegraphics[width = 0.5\textwidth]{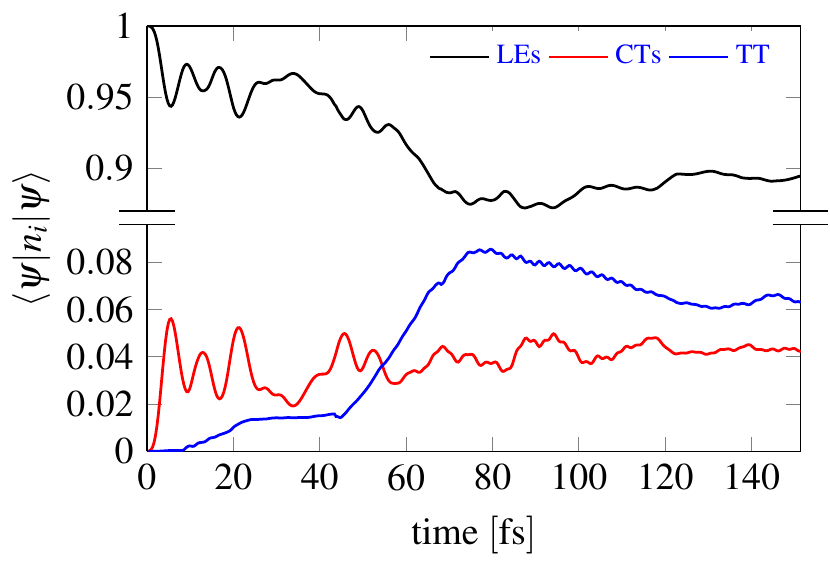} \fi
  }
  \caption{\label{fig:other_elec_occs}Electronic population dynamics for different initial states. \protect\cref{fig:loc_elec_population} corresponds to the localised initial state while \protect\cref{fig:dark_elec_population} is the dark one. Both were carried out with $E_\mathrm{TT} = \SI{2.9493}{\electronvolt}$.}
\end{figure}

We investigate the dependency of the excitonic population dynamics on the initial state.
To do so, we perform the time evolution with the excitation localised in one of the chromophores and with an anti\hyp symmetric excitation of both chromophones, as can be seen in~\cref{fig:other_elec_occs}.
It is to mention that the anti\hyp symmetric initial state is dark, i.e. it is not realisable in experiments from the ground state, due to transition rules.
We see that for \gls{le} initial state $\ket{\psi}(0) = \ket{\mathrm{LE}_1}$ the dynamics are similar to the bright initial state in the main text, i.e. there is also a delay from the \gls{ct}.
However the \gls{sf} yield is around half of the yield of the bright initial state.

As it was argued before, the mechanism behind the coherent \gls{sf} is the superposition of the \gls{le} and \gls{tt} state~\cite{Chan1541,doi:10.1146/annurev-physchem-040214-121235}.
This is the reason for the small yield of the localised state.
It is remarkable that a delocalisation over two chromophores gives a factor of two in the \gls{tt} yield.
This points into the direction of previous investigations~\cite{wang_isf_tetoli} emphasising the importance of spatial delocalisation.
Using the dark initial state we see that a much higher yield than for the system in the main text is reached.
After a plateau\hyp like region the value approaches $\sim\SI{10}{\percent}$.
However, there is a maximum after which the population begins to decay back mostly into the \gls{le} states.
\section{Excitonic \glsentrytext{rdm} analysis}\label{sec:rdmanalysis}
In the main text we already discussed the appearance of two different scales characterising the time evolution of the \gls{tt} occupancy.
In order to analyse the origin of these time scales we studied the \gls{rdm} of the excitonic system.
In general, the time evolution of the excitonic density operator $\hat \rho_\mathrm{ex}$ after tracing out the vibrational degrees of freedom can be expanded in its eigenbasis
\begin{align}
  \hat \rho_\mathrm{ex}(t) &= \sum_{i,j\in \mathcal S} \braket{i|\hat\rho_\mathrm{ex}(t)}{j} \ket{i}\bra{j} = \sum_n \lambda_n(t) \ket{\lambda_n(t)} \bra{\lambda_n(t)} \; \text{.}
\end{align}
Here, we explicitly exhibited the time\hyp dependence of the eigensystem and defined\\$\mathcal S = \left\{ \mathrm{\gls{tt}}, \mathrm{\gls{ct}}_2, \mathrm{\gls{ct}}_1,  \mathrm{\gls{le}}_2, \mathrm{\gls{le}}_1 \right\}$.
\begin{figure}[!h]
  \centering
  \iflocal \input{figures/elec_rdm} \else
  \includegraphics[width = \textwidth]{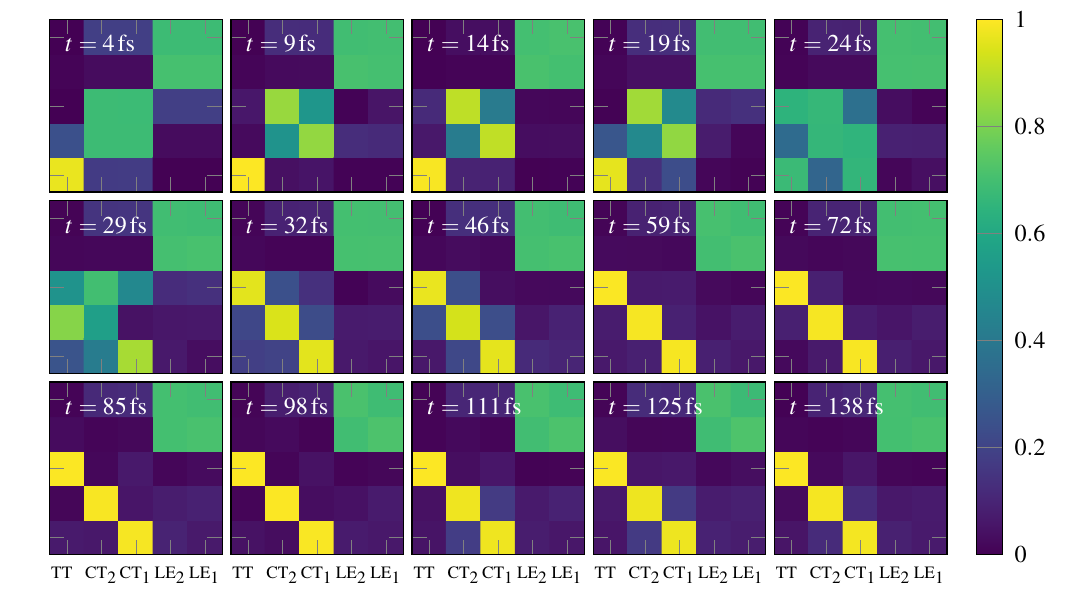} \fi
  \caption{\label{fig:elec_rdm} Time\hyp evolution of the eigenvectors $\lvert\braket{\lambda}{i}\rvert$ of the electronic \gls{rdm}. $\ket{\lambda}$ enumerates the \gls{rdm} eigenvalues while $\ket{i} \in \{\ket{\mathrm{TT}}, \ket{\mathrm{CT}_2}, \ket{\mathrm{CT}_1}, \ket{\mathrm{LE}_2}, \ket{\mathrm{LE}_1}\}$. Note that the eigenvalues are sorted by magnitude. We count them from top (largest) to bottom (smallest).}
\end{figure}
In~\cref{fig:elec_rdm} we show the absolute value of the decomposition of the eigenstates w.r.t. the excitionic states $\lvert\braket{\lambda}{i}\rvert$ at different stages during the time evolution.
We can clearly identify the timescale $t_{0} \approx \SI{30}{\femto\second}$.
At times $t<t_0$ the different eigenvalues of $\hat \rho_\mathrm{ex}$ are mainly coherent superpositions of the excitonic states.
The situation changes significantly at times $t>t_0$.
Here, the decomposition of the eigenstates separates into a block\hyp diagonal form. 
A mixing of excitonic states between the \gls{le}, \gls{ct} and \gls{tt} subsystems is then only subleading at the order of $\sim 10^{-1}$.
Ignoring the weak electronic coherences at $t>t_0$ this implies a decomposition
\begin{align}
  \hat \rho_\mathrm{ex}(t>t_0) &\approx \hat \rho_\mathrm{\gls{le}}(t) \oplus \hat \rho_\mathrm{\gls{ct}}(t) \oplus \hat \rho_\mathrm{\gls{tt}}(t)
  =
  \exp\left(-\left(\hat H_\mathrm{\gls{le}}(t) \oplus \hat H_\mathrm{\gls{ct}}(t) \oplus \hat H_\mathrm{\gls{tt}}(t)\right)t\right)/Z(t) \, .
\end{align}
In the last equation we introduced effective Hamiltonians governing the time evolution in the different excitonic subsystems:
\begin{align}
  \hat H_\mathrm{\gls{le}}(t) 
  &= 
  -\sum_{i,j=1,2} 
    \braket{\mathrm{\gls{le}}_i}{\lambda_4(t)} \log \lambda_4(t) \braket{\lambda_4(t)}{\mathrm{\gls{le}}_j} \, \ket{\mathrm{\gls{le}}_j} \bra{\mathrm{\gls{le}}_i} \notag \\
  &\phantom{=\sum_{i,j=1,2}}+
    \braket{\mathrm{\gls{le}}_i}{\lambda_5(t)} \log \lambda_5(t) \braket{\lambda_5(t)}{\mathrm{\gls{le}}_j} \, \ket{\mathrm{\gls{le}}_j}\bra{\mathrm{\gls{le}}_i} \\
    \hat H_\mathrm{\gls{ct}}(t) &= - \left(\ket{\mathrm{\gls{ct}}_1} \log \lambda_1(t) \bra{\mathrm{\gls{ct}}_1} + \ket{\mathrm{\gls{ct}}_2} \log \lambda_2(t) \bra{\mathrm{\gls{ct}}_2} \right) \\
    \hat H_\mathrm{\gls{tt}} (t) &= - \ket{\mathrm{\gls{tt}}} \log \lambda_3(t) \bra{\mathrm{\gls{tt}}}
\end{align}
Note that due to the normalisation of $\rho_\mathrm{ex}$ we have $\sum\limits_n \lambda_n = 1$.
In this decomposition of the excitonic density operator (at times $t>t_0$), the time evolution of the \glspl{ct} and \gls{tt} states is described by the dynamics of a purely mixed state without quantum mechanical coherences.
Furthermore, the dynamics of the \glspl{le} w.r.t. the remaining excitonic states is of the same type and coherent dynamic only happens within the \glspl{le}.
In order to investigate the origin of the found electronic coherence time $t_0$ we rewrite equation~\cref{eq:Hamiltonian} by means of a generalised Lang\hyp Firsov\hyp transformation~\cite{lang1963kinetic}.
For that purpose we define a transformation
\begin{align}
  \hat U &= \enumber^{-\hat S}\text{,} \quad \hat S = \sum_{ij,I} \lambda_{ij,I} \hat c^\dagger_i \hat c^\nodagger_j \left(\hat b^\dagger_I - \hat b^\nodagger_I\right) \;\text{,}
\end{align}
where $\lambda_{ij,I}$ are real coefficients.
These coefficients are now determined such that the exciton\hyp phonon couplings are cancelled in the transformed Hamiltonian $\enumber^{\hat S} \hat H \enumber^{-\hat S}$.
Expansion of the last expression yields:
\begin{align}
  \enumber^{\hat S} \hat H \enumber^{-\hat S} &= \hat H + \comm{\hat S}{\hat H} + \frac{1}{2} \comm{\hat S}{\comm{\hat S}{\hat H}} + \cdots \; \text{.}
\end{align}
After some algebra the commutator evaluates to
\begin{align}
  \comm{\hat S}{\hat H}
  &=
  \sum_{ij,I} \hat c^\dagger_i \hat c^\nodagger_j \left( A_{ij,I} - A_{ji,I} \right) \left(\hat b^\dagger_I - \hat b^\nodagger_I\right)
  -
  \sum_{ij,I} \lambda_{ij,I} \omega_I \hat c^\dagger_i \hat c^\nodagger_j \left( \hat b^\dagger_I + \hat b^\nodagger_I \right)
  \notag\\
  &\phantom{=}
  +
  \sum_{ij,IJ} \hat c^\dagger_i \hat c^\nodagger_j \left( B_{ij,IJ} - B_{ji,IJ} \right) \left(\hat b^\dagger_I + \hat b^\nodagger_I\right) \left(\hat b^\dagger_J - \hat b^\nodagger_J\right)
  -
  2 \sum_{ijkl} \Omega_{ijkl} \hat c^\dagger_i \hat c^\nodagger_j \hat c^\dagger_k \hat c^\nodagger_l \; \text{,}
\end{align}
where we defined $\lambda_{ij,I} = \frac{g_{ij,I}}{\omega_I}$ that generates the desired cancellation of the exciton\hyp phonon interaction term as well as
\begin{align}
  A_{ij,I} = \frac{1}{\omega_I} \sum_k g_{ik,I} V_{kj} \text{,}
  \quad
  B_{ij,IJ} = \frac{1}{\omega_I} \sum_k g_{ik,I} g_{kj,J}
  \quad \text{and} \quad
  \Omega_{ijkl} = \sum_I \frac{g_{ij,I} g_{kl,I}}{\omega_I} \; \text{.}
\end{align}
Furthermore $A_{ij,I}$ and $B_{ij,IJ}$ are symmetric matrices in the excitonic indices $i$ and $j$.
Thus the commutator of the generator with the Hamiltonian becomes
\begin{align}
  \comm{\hat S}{\hat H}
  &=
  -
  \sum_{ij,I} g_{ij,I} \hat c^\dagger_i \hat c^\nodagger_j \left( \hat b^\dagger_I + \hat b^\nodagger_I \right)
  -
  2 \sum_{ijkl} \Omega_{ijkl} \hat c^\dagger_i \hat c^\nodagger_j \hat c^\dagger_k \hat c^\nodagger_l \; \text{.}
\end{align}
In the following, higher order commutators including the quartic excitonic coupling term are neglected, since $\Omega_{ijkl}$ is subleading.
We thus obtain
\begin{align}
  \enumber^{\hat S} \hat H \enumber^{-\hat S}%
  &\approx
  \sum_{ij} V_{ij} \hat c^\dagger_i \hat c^\nodagger_j
  +
  \sum_I \omega_I \hat b^\dagger_i \hat b^\nodagger_I
  -
  2 \sum_{ijkl} \Omega_{ijkl} \hat c^\dagger_i \hat c^\nodagger_j \hat c^\dagger_k \hat c^\nodagger_l \; \text{.}
\end{align}
Transforming back into the exciton\hyp phonon basis we arrive at
\begin{align}
  \label{eq:lang-firsov-transformed-hamiltonian}
  \hat H
  &\approx
  \sum_{ij} V_{ij} \enumber^{-\hat S} \hat c^\dagger_i \hat c^\nodagger_j \enumber^{\hat S}
  +
  \sum_I \omega_I \enumber^{-\hat S} \hat b^\dagger_i \hat b^\nodagger_I \enumber^{\hat S}
  -
  2 \sum_{ijkl} \Omega_{ijkl} \enumber^{-\hat S} \hat c^\dagger_i \hat c^\nodagger_j \hat c^\dagger_k \hat c^\nodagger_l \enumber^{\hat S} \; \text{.}
\end{align}
Now we focus on the renormalisation of the exciton coupling matrix elements $V_{ij}$.
For that purpose we write the exponential in the eigenbasis of the excitonic part of the generator coefficients $\lambda_{ij,I} = \frac{g_{ij,I}}{\omega_I}$:
\begin{align}
  \exp\left(\sum_{I} \hat{\mathbf c}^\dagger \Lambda_{I} \hat{\mathbf c}^\nodagger \left(\hat b^\dagger_I - \hat b^\nodagger_I\right)\right) = \exp\left(\sum_I\hat{\mathbf f}^\dagger_I D_I \hat{\mathbf f}^\nodagger_I \left(\hat b^\dagger_I - \hat b^\nodagger_I\right) \right) \; \text{.}
\end{align}
In the last line we defined $\hat{\mathbf c}^\dagger = \left(\hat c^\dagger_\mathrm{LE1}, \hat c^\dagger_\mathrm{LE2}, \cdots \right)$, $D_I = R^\nodagger_I \Lambda_I R^\dagger_I$ with orthogonal matrices $R_I$ and rotated excitonic operators $\hat{\mathbf f}^\dagger_I = \hat{\mathbf c}^\dagger R^\dagger_I$.
Reinserting this expansion into the excitonic coupling term of~\cref{eq:lang-firsov-transformed-hamiltonian} yields
\begin{align}
   V_{ij} \enumber^{-\hat S} \hat c^\dagger_i \hat c^\nodagger_j \enumber^{\hat S}
   &=
   \exp\left(-\sum_{I} \hat{\mathbf f}^\dagger_I D_I \hat{\mathbf f}^\nodagger_I \left(b^\dagger_I - \hat b^\nodagger_I\right)\right) V_{ij} \hat c^\dagger_i \hat c^\nodagger_j \exp\left(\sum_{I} \hat{\mathbf f}^\dagger_I D_I \hat{\mathbf f}^\nodagger_I \left(\hat b^\dagger_I - \hat b^\nodagger_I\right)\right)
\end{align}
For each exciton\hyp phonon coupling matrix $\Lambda_I$ we computed the eigenvalues which we found to be strictly positive.
Additionally, each element of $\hat{\mathbf f}^\dagger_I \hat{\mathbf f}^\nodagger_I$ is between zero and one.
These observations motivate an expansion of the hopping elements of the transformed Hamiltonian to first order in the generator coefficients
\begin{align}
  V_{ij} \enumber^{-\hat S} \hat c^\dagger_i \hat c^\nodagger_j \enumber^{\hat S}
  &=
  V_{ij}
  \left(
    \hat c^\dagger_i \hat c^\nodagger_j
    -
      \hat c^\dagger_i \hat c^\nodagger_j
      \imagunit \sum_{kl}
        \sum_{I} \lambda_{kl,I} \hat c^\dagger_k \hat c^\nodagger_l \imagunit(\hat b^\dagger_I - \hat b^\nodagger_I)
      +
      \imagunit \sum_{kl}
        \sum_{I} \lambda^*_{kl,I} \hat c^\dagger_k \hat c^\nodagger_l \imagunit(\hat b^\dagger_I - \hat b^\nodagger_I)
      \hat c^\dagger_i \hat c^\nodagger_j
    +
    \cdots
  \right) \notag \\
  &=
  V_{ij}
  \left(
    \hat c^\dagger_i \hat c^\nodagger_j
    -
      \imagunit \sum_{l} \hat c^\dagger_i \hat c^\nodagger_l
        \underbrace{\sum_{I} \lambda_{jl,I} \imagunit(\hat b^\dagger_I - \hat b^\nodagger_I)}_{= \Delta \hat U_{jl}}
      +
      \text{h.c.}
    +
    \cdots
  \right)
  \; ,
\end{align}
where we explicitly used the fact that we are working in a single\hyp exciton subspace.
Here, we also introduced the hopping energy renormalisation $\Delta \hat U_{jl}$, which is referred to in~\cref{fig:polaron-sketch} in the main text.
As a consequence, an enhancement (positive overall exponent) or suppression (negative overall exponent) of the hoppings can be identified by evaluating the momentum expectation values $\propto \imagunit(\hat b^\dagger_I - \hat b^\nodagger_I)$.
\begin{figure}[!h]
  \centering
  \iflocal \input{figures/summed_polaron_coeffs} \else
  \includegraphics{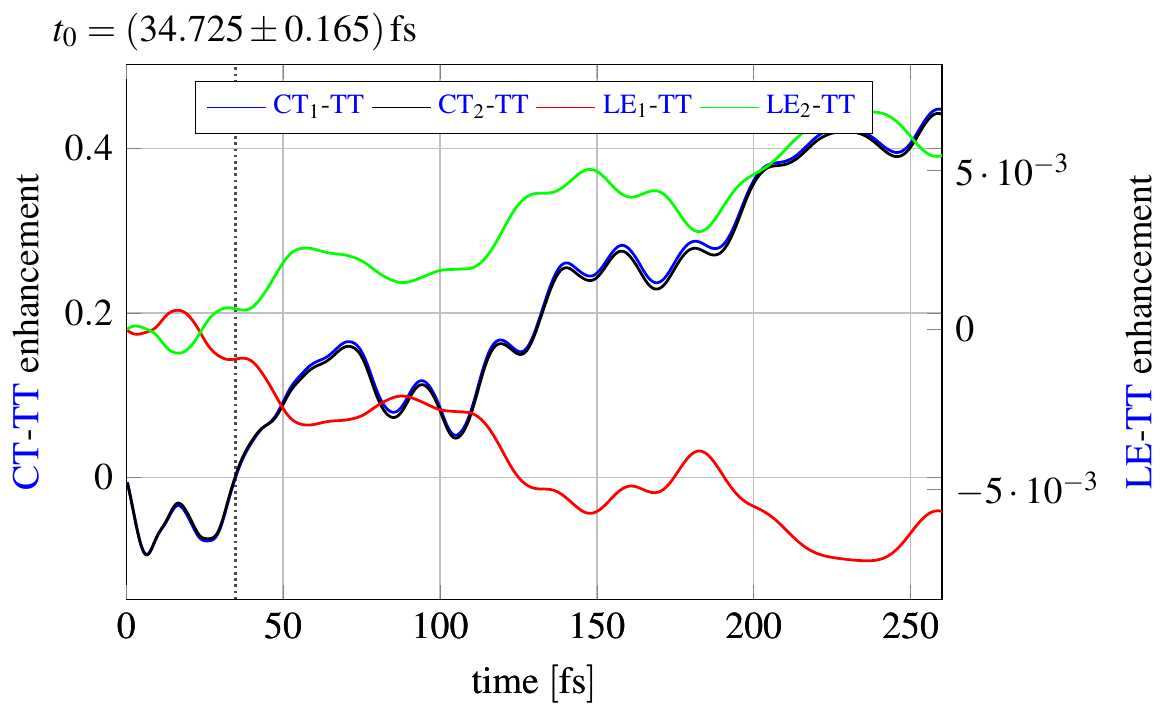} \fi
  \caption{\label{fig:summed-polaron-coeffs}Summed up transformation matrix elements corresponding to transitions into the \gls{tt} state. The sign change indicates a transition from enhanced to suppressed hoppings.}
\end{figure}
In~\cref{fig:summed-polaron-coeffs} we show the expectation value of the hopping matrix renormalisation corresponding to transitions into the \gls{tt} state
\begin{align}
	\expval{\sum\limits_I \lambda_{\text{\gls{tt}}j,I} \imagunit (\hat b^\dagger_I - \hat b^\nodagger_I)}{\psi(t)} \quad \text{with $j=\text{\gls{le}}_1,\text{\gls{le}}_2,\text{\gls{ct}}_1,\text{\gls{ct}}_2$}
\end{align}
We find a change of sign at a simulation time $t_0 = \SI{34.725}{\femto\second}$.
Note that for $t < t_0$ we observe a negative sign in the dominating matrix elements corresponding to hopping from the \gls{ct} into the \gls{tt} state indicating an enhancement of the transition amplitude.
For $t > t_0$ the situation changes into the opposite and hoppings into the \gls{tt} state are suppressed.
Importantly, we also find a nearly vanishing renormalisation for the direct path \gls{le} $\longrightarrow$ \gls{tt}.
This emphasises the significant role played by the super\hyp exchange path in the \gls{tt} production during the coherent dynamics.

%
%
\end{document}